\documentclass[aoas]{imsart}

\RequirePackage{amsthm,amsmath,amsfonts,amssymb}
\RequirePackage[authoryear]{natbib}
\RequirePackage[colorlinks,citecolor=blue,urlcolor=blue]{hyperref}
\RequirePackage{graphicx}
\RequirePackage{float, subfig}
\usepackage{dcolumn, multirow}
\usepackage{bm}
\usepackage{tabularx}
\newcolumntype{L}{D{.}{.}{2,4}}
\newcolumntype{Y}{>{\centering\arraybackslash}X}

\startlocaldefs
\defcitealias{WPP2022}{United Nations (2022)} 

\endlocaldefs

\begin{document}
\graphicspath{{Figures/}}

\begin{frontmatter}
\title{Multiple Imputation of Hierarchical Nonlinear Time Series Data with an Application to School Enrollment Data}
\runtitle{Multiple Imputation of Hierarchical Nonlinear Time Series Data}

\begin{aug}
\author[A]{\fnms{Daphne H.} \snm{Liu}\ead[label=e1]{liu.daphneh@gmail.com}\orcid{0000-0002-1541-4279}}, 
\and
\author[B]{\fnms{Adrian E.} \snm{Raftery}\ead[label=e2]{raftery@uw.edu}\orcid{0000-0002-6589-301X}}
\address[A]{Department of Statistics, University of Washington
\printead{e1}}

\address[B]{Departments of Statistics and Sociology,
University of Washington,
\printead{e2}}
\end{aug}

\begin{abstract}
International comparisons of hierarchical time series data sets based on survey data, such as annual country-level estimates of school enrollment rates, can suffer from large amounts of missing data due to differing coverage of surveys across countries and across times. 
A popular approach to handling missing data in these settings is through multiple imputation, which can be especially effective when there is an auxiliary variable that is strongly predictive of and has a smaller amount of missing data than the variable of interest. 
However, standard methods for multiple imputation of hierarchical time series data can perform poorly when the auxiliary variable and the variable of interest have a nonlinear relationship.
Performance can also suffer if the multiple imputations are used to estimate an analysis model that makes different assumptions about the data compared to the imputation model, leading to uncongeniality between analysis and imputation models. 
We propose a Bayesian method for multiple imputation of hierarchical nonlinear time series data that uses a sequential decomposition of the joint distribution and incorporates smoothing splines to account for nonlinear relationships between variables.
We compare the proposed method with existing multiple imputation methods through a simulation study and an application to secondary school enrollment data.
We find that the proposed method can lead to substantial performance increases for estimation of parameters in uncongenial analysis models and for prediction of individual missing values. 
\end{abstract}

\begin{keyword}
\kwd{Multiple imputation, missing data, Bayesian hierarchical model, time series, school enrollment}
\end{keyword}

\end{frontmatter}


\section{Introduction}
Missing values within hierarchical time series data sets are a common occurrence in social science data, particularly for comparisons across countries and across times that rely on survey data. 
International surveys may only be conducted in selected years and coverage of countries may differ from year to year, while country-level surveys and censuses may be conducted annually in some countries but may only occur sporadically in others.
The presence of missing data can be compounded in settings that require the compilation of data from multiple different sources.
One example where this occurs is for school enrollment data, where the UNESCO Institute for Statistics compiles survey and administrative data from countries around the world to create annual, internationally comparable estimates of school enrollment rates.
The country-specific time series for school enrollment rates have a hierarchical structure where countries are nested in the world.
The differing availability of survey and administrative data across countries and years leads to a large number of country-years in the UNESCO database with missing data.

For hierarchical time series data arising from survey data, there can be multiple variables that measure similar underlying quantities but have differing amounts and patterns of missing data.
This is the case for school enrollment data, where two commonly reported measures of school enrollment rates are the Net Enrollment Rate (NER) and the Gross Enrollment Ratio (GER).
NER is the ratio of children of official secondary school age who are enrolled in secondary school to the population of official secondary school age children, while GER is the ratio of total enrollment in secondary school, regardless of age, to the population of official secondary school age children. 
NER and GER both aim to measure the same underlying quantity of school enrollment.
NER can be thought of as a more refined measure of school enrollment that incorporates information about the age of children who are enrolled in school, while GER is a coarser measure of school enrollment that only requires information about the number of children enrolled.
NER is more difficult to measure than GER and has a larger amount of missing data. 

More generally, this type of missing data can arise if a survey is designed to collect information on a refined measure of the quantity of interest but includes the option to collect information on a coarse measure of the quantity of interest if the respondent does not provide enough information for the refined measure.
For example, a survey on school enrollment might ask for information on the number and ages of children enrolled in school to calculate NER but include the option for respondents to provide only the number of children enrolled to calculate GER.
In surveys that do not collect the coarse measure directly, there may exist auxiliary information from external sources, such as administrative records, that can be used as a coarse measure of the quantity of interest.
This type of missing data can also occur by design in longitudinal panels to reduce respondent burden.
Panel members may be asked to provide basic information for the coarse measure in every cycle of the panel, but may only be asked to provide detailed information for the refined measure in selected cycles. 
In all of these examples, the refined measure of the quantity of interest tends to have a greater amount of missing data than the coarse measure by nature of being more difficult to measure. 
If the variable of interest for subsequent analysis is the refined measure, researchers might be interested in how to best leverage information from the coarse measure to impute the variable of interest using a multiple imputation procedure. 

Multiple imputation, first developed by \cite{Rubin1978, Rubin1987}, is a widely used approach for handling missing data.
In multiple imputation, $M > 1$ imputed values for missing observations are sampled from the posterior predictive distribution of the missing data given the observed data. 
The $M$ imputed values result in $M$ completed data sets, each of which consists of the observed data and one set of imputed values for the missing observations. 
The completed data sets can each be analyzed separately using complete data methods and the results of the analyses can be combined into one final, pooled result using combining rules from \cite{Rubin1987} that account for both within-imputation variation and between-imputation variation.
Multiple imputation approaches generally assume data is Missing At Random (MAR) as defined by \cite{Rubin1976}.
For variable $\mathbf{X}$, MAR occurs when the probability of being missing depends only on the observed portion of $\mathbf{X}$.
A special case of MAR occurs when data is Missing Completely At Random (MCAR) and the probability of being missing does not depend on $\mathbf{X}$.
Data can also be Missing Not At Random (MNAR), which occurs when the probability of being missing depends on the missing portion of $\mathbf{X}$.

We focus on the setting where the imputation of missing values and the analysis of imputed values are conducted independently.  
The development of multiple imputation originated in this setting, where \cite{Rubin1977, Rubin1978} proposed the multiple imputation framework as a way to provide imputed values for missing responses in public-use releases of large data sets from sample surveys. 
An appealing feature of multiple imputation for practitioners is the ability to use the same imputed data set to conduct many different analyses (\cite{Schafer1997, Rubin1987}).
However, using the same multiply imputed data for different analyses can lead to the analysis model being uncongenial to the imputation model in the sense of \cite{Meng1994}.
An imputation model and an analysis model are congenial if there exists a Bayesian model such that (i) the posterior mean and variance from the Bayesian model for the parameter of interest are asymptotically the same as the mean and variance estimates from the analysis model in both the complete and incomplete data settings and (ii) the posterior predictive distribution of the missing data given the observed data derived from the Bayesian model is identical to the imputation model \citep{Meng1994}. 
When these two conditions are not met, the imputation and analysis models are uncongenial, and theoretical properties of the multiply imputed estimates created using the simple combining rules of \citet{Rubin1987} may not be guaranteed (\cite{Meng1994, Rubin1996, XieMeng2017}). 
Rubin's multiply imputed variance estimator is not guaranteed to be asymptotically unbiased for the true repeated sampling variance of the multiply imputed point estimator under uncongeniality, leading to estimated confidence intervals with less than nominal coverage. 
Uncongeniality of the analysis and imputation models can be a regular occurrence in practice, particularly when the researchers that collect the survey data create imputed values for publication that are then used in analyses by external researchers.
The external researchers may not have access to the same information or resources needed to create imputations of their own, for example if the variables used in the imputation model are not publicly available.
Specifically, we consider the scenario where imputation of a variable $\mathbf{Y}$ for public release is conducted by leveraging the relationship between $\mathbf{Y}$ and an auxiliary variable $\mathbf{X}$, where $\mathbf{X}$ (and imputed values of $\mathbf{X}$) cannot be made publicly available. 
The multiply imputed values of $\mathbf{Y}$ are subsequently used by an external researcher in an analysis with an external outcome variable, $\mathbf{Z}$, which is unknown to the imputer and is not accounted for in the imputation process. 
In this scenario, the analysis model used by the external researcher is not guaranteed to be congenial to the imputation model. 
The ability of a multiple imputation method to perform well for uncongenial analyses is thus of interest for practitioners.

For hierarchical data, multiple imputation approaches that do not explicitly account for the hierarchical structure of the data can lead to biased results in downstream analyses (\cite{TaljaardDonnerKlar2008, EndersMistlerKeller2016, LudtkeRobitzschGrund2017}).
Many approaches specifically for multiple imputation of hierarchical time series data have been developed (e.g. \cite{LiuTaylorBelin2000, HeYucelRaghunathan2011, Speidel2018, EndersDuKeller2020, GrundLudtkeRobitzsch2021}; among others), with two of the most widely used approaches for social science data being Amelia II and multilevel extensions of Multiple Imputation by Chained Equations (MICE). 
Amelia, originally developed by \cite{King2001} and extended as Amelia II by \cite{HonakerKing2010}, is a multiple imputation method designed specifically for hierarchical time series data.
Amelia is based on the joint modeling approach to multiple imputation, where  imputed values are sampled from a joint distribution for all variables with missing data. 
MICE is a multiple imputation method developed by \cite{vanBuurenOudshoorn2011} that uses the fully conditional specification (FCS) approach to multiple imputation. 
Rather than explicitly specifying a joint imputation model, the FCS algorithm iteratively samples from univariate conditional imputation models for all variables with missing data until convergence is reached.
Several methods that account for hierarchical data structures have been implemented within the MICE framework, including the linear mixed effects method developed by \cite{SchaferYucel2002}. 

The most commonly used methods for multiple imputation of hierarchical time series data assume that the variables in the imputation model have a linear relationship.
In settings where variables have a strong nonlinear relationship and transformation to approximate linearity is not possible, these methods are  misspecified.
Failing to account for the nonlinear relationship can lead to imputed relationships between variables that are implausible based on substantive knowledge and biased estimates in analyses using the multiply imputed data. 
Nonlinear relationships in hierarchical time series data can occur regularly in the social sciences, for example if rates that aim to measure the same quantity are calculated for related populations.
This is the case for school enrollment rates, where NER and GER measure school enrollment for slightly different populations of children. 
Another example arises for contraceptive prevalence rates, where country-level estimates of contraceptive prevalence for married and unmarried women have a nonlinear relationship when pooled across countries and times. 
Measurement of contraceptive use is more difficult for unmarried women than for married women, leading to a larger amount of missing data for unmarried women. 
A multiple imputation method for hierarchical time series data that can account for this type of nonlinear relationship could be of great utility for such applications. 
Existing multiple imputation methods that address more general nonlinear settings include methods based on classification and regression trees (\cite{BurgetteReiter2010}), functional data analysis (\cite{HeYucelRaghunathan2011}), B-splines (\cite{Mbougua2013}), and generalized additive models for location, scale, and shape (\cite{deJong2016}). 
We build upon this existing literature to develop a multiple imputation method for continuous hierarchical time series data that can account for a nonlinear relationship between variables that represent refined and coarse measures of the underlying quantity of interest using smoothing splines.
We refer to this method as MINTS for Multiple Imputation of hierarchical Nonlinear Time Series data.
We focus on the bivariate setting, where the refined measure is the variable for which imputations are desired, and the coarse measure is an auxiliary variable that is easier to measure and has a nonlinear relationship with the refined measure.

This paper is structured as follows.
In Section 2, we describe the motivating case study of secondary school enrollment data in further detail. 
Section 3 describes the proposed multiple imputation method.
In Section 4, we conduct a simulation study to evaluate the out-of-sample validation performance of the proposed multiple imputation method.
We compare the performance of the proposed multiple imputation method with several existing multiple imputation methods for estimation of parameters in analysis models that are uncongenial to the imputation model.
In Section 5, we conduct two out-of-sample validation exercises using the motivating data set on secondary school enrollment rates, where we evaluate predictive performance for estimation of individual missing values and estimation of parameters in uncongenial analysis models.
We also create multiple imputations for the full school enrollment data set and make available 40 multiple imputations for NER created using MINTS.
Section 6 includes further discussion and comparisons of the proposed multiple imputation method with existing methods.
Finally, we summarize the findings of this paper in Section 7.

\section{Motivating Case Study: Secondary School Enrollment Rates} \label{sec:data}
The UNESCO Institute for Statistics collects internationally comparable data on education indicators on an annual basis for all countries of the world, based largely on survey and administrative data (\cite{UNESCO}).
The World Bank combines this education data with population data from the United Nations Population Division to create estimates of two types of enrollment rates: the Net Enrollment Rate (NER) and the Gross Enrollment Ratio (GER) (\cite{WorldBank}). 
We focus on secondary school enrollment.
NER is the ratio of children of official secondary school age who are enrolled in secondary school to the population of official secondary school age children.
NER is bounded between 0\% and 100\% and both the numerator and denominator reflect children of official secondary school age.
GER is the ratio of total enrollment in secondary school, regardless of age, to the population of official secondary school age children. 
The numerator and denominator for GER potentially represent different populations, where children who are not of official secondary school age can be counted in the numerator but not in the denominator. 
Thus, GER can be greater than 100\% if children who are enrolled in school are not of official school age.
The two measures of enrollment have a strong nonlinear relationship and are subject to the boundary NER $\leq$ GER. 

For substantive analyses, one measure of enrollment may be preferred over the other.
NER can be thought of as a demographic rate, where the numerator counts the number of enrollments for the population of children of official secondary school age in a given year and the denominator counts the person-years lived in that population for the given year. 
NER can thus be preferable over GER for demographic analyses. 
However, historical time series of NER tend to have more missing values than time series of GER, as measurement of NER is more difficult than measurement of GER due to requiring knowledge of the age distribution for children enrolled in school. 
Measurement of GER is comparatively easy, as GER can be calculated using only the number of children who are currently enrolled. 
For school systems that do not have robust recordkeeping systems and countries that do not have good vital registration systems, knowledge of the age of all enrolled children can be difficult to obtain. 
The greater availability of estimates of GER and the strong relationship between NER and GER motivates the desire to impute missing values of NER using the relationship between NER and GER.

We obtain estimates of secondary school enrollment rates for both genders combined from \cite{WorldBank},\footnote{Downloaded on August 5, 2021} where the definition of secondary school used for each country is based on the International Standard Classification of Education.
After excluding all countries and years in the World Bank data base with no observations for either NER or GER, the resulting data set includes 202 countries and 51 years spanning 1970 to 2020 for a total of 10,302 country-year combinations.
The overall rate of missingness is about 73.0\% for NER and about 37.6\% for GER.
Within countries, the rate of missingness for NER ranges from about 13.7\% missing in Malta to 100\% missing in 14 countries. 
For GER, the rate of missingness within countries ranges from about 2.0\% missing in Peru to about 98.0\% missing in Cura\c{c}ao.

\begin{figure}[!htb]
	\centering
	\includegraphics[width = 0.5\textwidth]{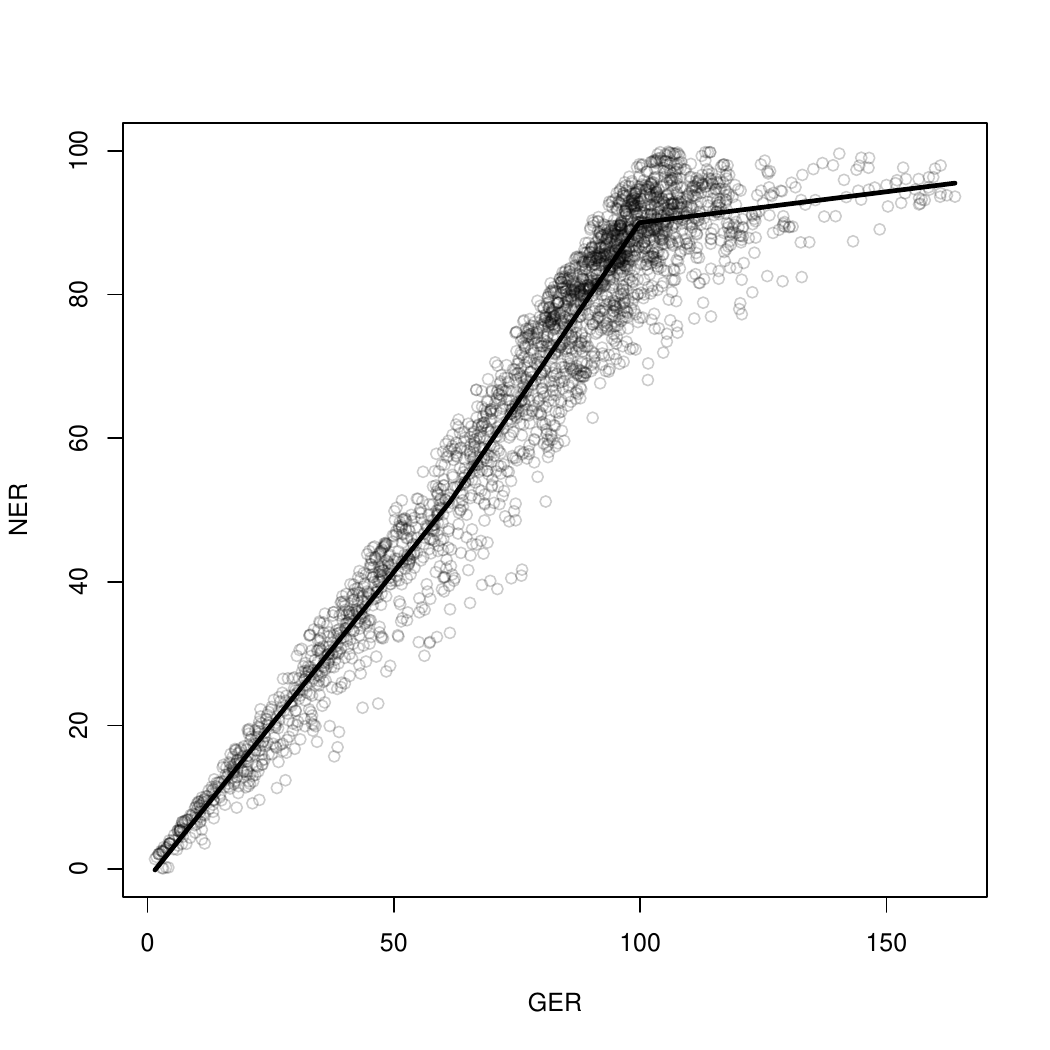}
	\caption{Scatter plot of complete cases for NER and GER from secondary enrollment data set superimposed with the B-spline of degree 1 fit using A-splines.} \label{fig:f}
\end{figure}

Figure \ref{fig:f} shows a scatter plot of the complete cases for NER and GER. 
The superimposed line illustrates the B-spline of degree 1 fit to the complete cases using the A-splines methodology described in Section \ref{sec:aspline}. 
There is a nonlinear relationship between NER and GER, with a shift in trend occurring around GER $= 100$. 
The variation of NER about the fitted spline also appears to vary with GER, with smaller variability at the lowest levels of GER and larger variability around GER $= 100$. 

\begin{figure}[!htb]
	\centering
	\includegraphics[width = 0.65\textwidth]{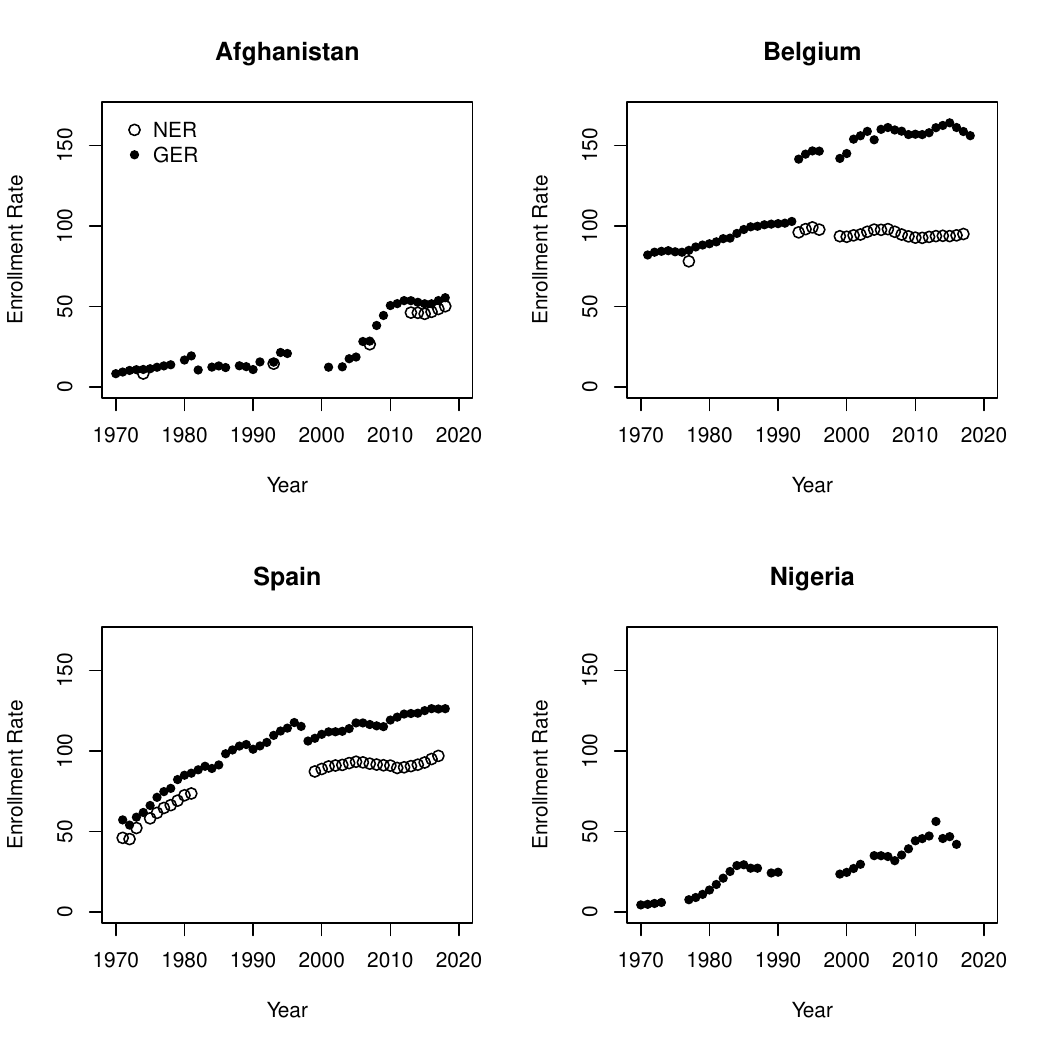}
	\caption{Observed values of NER and GER for selected countries from secondary enrollment data set. Open and solid circles indicate observed values for NER and GER, respectively.} \label{fig:example_countries}
\end{figure}

Time series of NER and GER are shown in Figure \ref{fig:example_countries} for Afghanistan, Belgium, Spain, and Nigeria.
Afghanistan is an example of the most common type of pattern seen for individual countries, where there is a larger number of observed values for GER compared to NER.
Only one country, Brazil, has the opposite pattern with one more observed value of GER than observed values of NER.
Belgium and Spain are two examples of countries where the nonlinear relationship between NER and GER is visible in the time series for the individual country. 
Both Belgium and Spain have relatively few missing values for GER, but have large stretches of time with no observations for NER.
Finally, Nigeria is an example of a country that has some observed values for GER but has no observed values for NER. 
There are 14 countries in the school enrollment data set that, like Nigeria, have at least one observation for GER but no observations for NER. 

\section{Methods}
\subsection{Notation}
We consider hierarchical time series data where the clustering variable is country and the time variable is year. 
Let $X_{c,t}$ denote the auxiliary variable and let $Y_{c,t}$ denote the variable of interest for country $c$ and year $t$, where $c \in 1, ..., C$ and $t \in 1, ..., T$.
The vector of $X_{c,t}$ for all countries at year $t$ is denoted by $\mathbf{X}_t = \left[X_{1,t}, X_{2,t}, ..., X_{C,t} \right]$.
Similarly, the vector of $Y_{c,t}$ for all countries at year $t$ is denoted by $\mathbf{Y}_t = \left[Y_{1,t}, Y_{2,t}, ..., Y_{C,t} \right]$.
The $C$ by $T$ matrix of all $X_{c,t}$ is $\mathbf{X} = \left[ \mathbf{X}_1, \mathbf{X}_2, ..., \mathbf{X}_T \right]$ and the $C$ by $T$ matrix of all $Y_{c,t}$ is $\mathbf{Y} = \left[ \mathbf{Y}_1, \mathbf{Y}_2, ..., \mathbf{Y}_T \right]$.

Let the matrices $\mathbf{R}^X$ and $\mathbf{R}^Y$ denote the response matrices for $\mathbf{X}$ and $\mathbf{Y}$, respectively. 
If an element $(c,t)$ is observed, then the corresponding element of the response matrix equals 1. 
For example, if $X_{c,t}$ is observed, then $R_{c,t}^X = 1$. 
If $X_{c,t}$ is missing, then $R_{c,t}^X = 0$. 
The matrices $\mathbf{X}$ and $\mathbf{Y}$ can also be written in terms of their observed and missing portions.
For example, $\mathbf{X} = [\mathbf{X}_{mis}, \mathbf{X}_{obs}]$ where the observed portion of $\mathbf{X}$ is denoted $\mathbf{X}_{obs}$ and the unobserved portion of $\mathbf{X}$ is denoted $\mathbf{X}_{mis}$.

For the school enrollment data, the auxiliary variable $\mathbf{X}$ is GER and the variable of interest $\mathbf{Y}$ is NER. 
The enrollment data includes a total of $C = 202$ countries, where the assignment of countries to indices $c \in 1, ..., C$ was done alphabetically by country name.
There are $T= 51$ years in the enrollment data set, where $t = 1$ corresponds to the year 1970 and $t = 51$ corresponds to the year 2020. 

\subsection{Model}
\subsubsection{Assumptions}
We assume the missing data mechanism is ignorable, i.e., the missing data mechanism is Missing At Random (MAR, defined in Section \ref{sec:mechanisms}) and the parameters of the complete data model for ($\mathbf{X}, \mathbf{Y}$) are distinct and a priori independent from the parameters of the missing data model governing the response matrices $\mathbf{R}^X$ and  $\mathbf{R}^Y$ (\cite{Rubin1976, Schafer1997, LittleRubin2002}). 
The assumption of ignorability allows for imputation without specification of a model for the missing data mechanism.
Using $\mathbf{X}$  as an example, imputed values are said to be Bayesianly proper in the sense of \cite{Schafer1997} if they are independently drawn from the posterior distribution $p(\mathbf{X}_{mis} | \mathbf{X}_{obs}, \mathbf{R}^X)$. 
When the missing data mechanism is ignorable, proper imputations can instead be drawn from the posterior distribution 
$p(\mathbf{X}_{mis} | \mathbf{X}_{obs}) = \int p(\mathbf{X}_{mis} | \mathbf{X}_{obs}, \theta) p(\theta | \mathbf{X}_{obs}) d \theta$
where $\theta$ is the vector of parameters for the complete data. 
The assumption of ignorability is ubiquitous for general-purpose imputation models, but in practice researchers cannot know if this assumption is met. 
We conducted sensitivity analyses in Sections \ref{sec:simulation} and \ref{sec:enroll} to evaluate how well the MINTS method performs when the ignorability assumption is violated.
	
Finally, we assume that the auxiliary variable is observed at least once for each country for which we are interested in imputing the variable of interest.
That is, both $\mathbf{X}$ and $\mathbf{Y}$ can contain missing values for any country-year, but we assume that each country has at least one year in which $X_{c,t}$ is observed. 
This reflects the motivating setting for the proposed imputation method, where the auxiliary variable is more likely to be observed in each country than the variable of interest due to being easier to measure.

\subsubsection{Sequential Decomposition of Joint Model}
The MINTS method uses a variation of the joint modeling approach to multiple imputation.
In the usual joint modeling approach, the complete data $(\mathbf{X}, \mathbf{Y})$ is assumed to follow a multivariate joint distribution and imputed values are sampled from the joint posterior predictive distribution of the missing data given the observed data.
One downside to this approach is the difficulty of specifying a joint distribution for all variables used in the imputation model.
A common general-purpose choice for the joint distribution is the multivariate normal distribution, which has been found to still perform reasonably well for imputation even when the assumption of normality is violated (\cite{Schafer1997, SchaferOlsen1998}).
However, if there are nonlinear relationships between variables in the imputation model, joint modeling assuming multivariate normality is unable to incorporate those relationships.

A more flexible approach to multiple imputation is the fully conditional specification (FCS) approach, also known as imputation via chained equations (\cite{vanBuuren2006}) and sequential regression multivariate imputation (\cite{Raghunathan2001}), among other names.
FCS does not explicitly specify a joint distribution for the variables with missing data. 
Instead, univariate conditional distributions are specified for each variable with missing data given all other variables in the imputation model. 
Missing values are imputed by iteratively sampling from the univariate conditional distributions until convergence for all imputation model parameters is reached. 
The conditional distributions can take any form and can accommodate nonlinear relationships between variables. 
However, this level of flexibility can lead to cases where the univariate conditional distributions are not compatible. 
Although simulation studies show that FCS can result in reasonable imputations even when the conditional distributions are not compatible (e.g. \cite{vanBuuren2006}), ultimately FCS does not guarantee that the iterative conditional algorithm will converge to a proper joint distribution (\cite{LiYuRubin2012}).

We use an alternative to the usual joint modeling approach that combines desirable features of joint modeling and FCS through a sequential decomposition of the joint model.
The sequential decomposition approach for joint modeling was first proposed by \cite{LipsitzIbrahim1996} and has been extended by \cite{Ibrahim1999, Ibrahim2002, LeeMitra2016, Xu2016, Ludtke2020}; among others. 
The joint distribution of the variables with missing data is decomposed into a sequence of univariate conditional distributions. 
One possible decomposition of the joint distribution for $(\mathbf{X}, \mathbf{Y})$ is
\begin{align} \label{eqn:decomp}
p(\mathbf{X}, \mathbf{Y} | \bm{\theta}) =& p(\mathbf{Y} | \mathbf{X}, \bm{\theta}_Y) p(\mathbf{X} | \bm{\theta}_X).
\end{align}
The vector of parameters for the joint distribution is $\bm{\theta}$, the vector of parameters for the conditional distribution of $\mathbf{Y} | \mathbf{X}$ is $\bm{\theta}_Y$, and the vector of parameters for the distribution of $\mathbf{X}$ is $\bm{\theta}_X$. 
Unlike in FCS, this sequence of univariate conditional distributions is guaranteed to correspond to a well-defined joint distribution by construction.

The sequential decomposition approach allows for greater flexibility compared to joint modeling, but is not quite as flexible as FCS due to the additional restriction of requiring an ordering for the conditional distributions.
The choice of ordering can have a substantial impact on the performance of the imputation model due to the risk of conditional distributions being incorrectly specified.
We follow standard guidelines proposed by \cite{RubinSchafer1990} and choose the ordering based on the percentage of missing values.
For the ordering in Equation \ref{eqn:decomp}, the variable of interest $\mathbf{Y}$ has a larger amount of missing data compared to the auxiliary variable $\mathbf{X}$.

\subsubsection{Model Specification} \label{sec:MINTS_model}
The joint distribution of $(\mathbf{Y}, \mathbf{X})$ is decomposed sequentially following Equation \ref{eqn:decomp} as the product of the distribution of $\mathbf{Y} | \mathbf{X}$ and the distribution of $\mathbf{X}$. 
The distribution of $\mathbf{X}$ is further decomposed as
\begin{align*}
p(\mathbf{X} | \bm{\theta}_X) =& \Bigg(\prod_{t = 1}^T p(\mathbf{X}_t | \mathbf{X}_{t-1}, \bm{\theta}_X)\Bigg) p(\mathbf{X}_0 | \bm{\theta}_X),
\end{align*}
where $\mathbf{X}_0$ is a parameter that represents the vector of $X_{c,0}$ for all countries $c$ at the unobserved year $t = 0$. 
Similarly, the conditional distribution of $\mathbf{Y} | \mathbf{X}$ is further decomposed as
\begin{align*}
p(\mathbf{Y} | \mathbf{X}, \bm{\theta}_Y) =& \left(\prod_{t = 1}^T p(\mathbf{Y}_t | \mathbf{Y}_{t-1}, \mathbf{X}_t, \bm{\theta}_Y)\right) p(\mathbf{Y}_0 | \mathbf{X}_0, \bm{\theta}_Y),
\end{align*}
where $\mathbf{Y}_0$ is a parameter that represents the vector of $Y_{c,0}$ for all countries $c$ at the unobserved year $t = 0$.

The distribution of $\mathbf{X}_t | \mathbf{X}_{t-1}, \bm{\theta}_X$ is modeled as a random walk with a country-specific drift term $\gamma_c$. 
For country $c$ and $t \in 1, ..., T$, 
\begin{align} \label{eqn:X_model}
X_{c,t} | X_{c,t-1}, \bm{\theta}_X \sim& \, TN_{[X_{low}, X_{up}]}(X_{c,t-1} + \gamma_c, \, \sigma_X^2), \\
\gamma_c \sim& N(\mu_{drift}, \sigma_{drift}^2), \nonumber
\end{align}
where $TN_{[X_{low}, X_{up}]}$ refers to the truncated normal distribution with lower bound $X_{low}$ and upper bound $X_{up}$.
These boundaries should be specified based on substantive knowledge about the range of $\mathbf{X}$ and can vary with $(c,t)$, but the dependency is suppressed in the notation. 
The country-specific drift terms $\gamma_c$ follow a world-level distribution where $\mu_{drift}$ and $\sigma_{drift}^2$ are world-level hyperparameters.

The conditional distribution of $\mathbf{Y}_t | \mathbf{Y}_{t-1}, \mathbf{X}_t, \bm{\theta}_Y$ is modeled with a country-specific intercept $\alpha_c$, a nonlinear function $f$ of $\mathbf{X}$ with coefficient $\beta$, and an AR(1) term with autoregressive parameter $\rho$. 
The variance model for $\mathbf{Y}_t | \mathbf{Y}_{t-1}, \mathbf{X}_t, \bm{\theta}_Y$ models heteroscedasticity as a function $h$ of $\mathbf{X}$. 
For country $c$ and $t \in 1, ... ,T$, 
\begin{align}  \label{eqn:Y_model}
Y_{c,t} | Y_{c,t-1}, X_{c,t}, \bm{\theta}_Y \sim& \, TN_{[Y_{low}, Y_{up}]}\left(\alpha_c + \beta \, f(X_{c,t}) + \rho \, Y_{c,t-1} , \, \sigma_{Y}^2  \, h(X_{c,t})\right ), \\
\alpha_c \sim& N(\mu_0, \sigma_0^2), \nonumber
\end{align}
where $TN_{[Y_{low}, Y_{up}]}$ refers to the truncated normal distribution with lower bound $Y_{low}$ and upper bound $Y_{up}$.
These boundaries should be specified based on substantive knowledge about the range of $\mathbf{Y}$ and can vary with $(c,t)$, but the dependency is suppressed in the notation. 
The country-specific intercepts $\alpha_c$ follow a world-level distribution where $\mu_0$ and $\sigma_0^2$ are world-level hyperparameters. 

\subsubsection{Prior Distributions} \label{sec:priors}
The prior distributions are specified as
    \begin{align*}
    \sigma_X^2 \sim& InvGamma(2, \delta_X), \\
    \mu_{drift} \sim& N(\nu_{drift}, \zeta_{drift}^2), \\
    \sigma_{drift^2} \sim& InvGamma(2, \delta_{drift}),  \\
    \sigma_{Y}^2 \sim& InvGamma(2, \delta_Y), \\
    %
    \mu_0 \sim& N(0, \zeta_0^2), \\
    \sigma_0^2 \sim& InvGamma(2, \delta_0), \\
    \beta \sim& N(0, 1), \\
    \rho \sim& U(0, 1), 
    \end{align*}
where the hyperparameters $\delta_X, \nu_{drift}, \zeta_{drift}, \delta_{drift}, \delta_Y, \zeta_0, \text{and } \delta_0$ are control parameters that are used to adjust the prior distributions to the appropriate scale for the data.

For all $c$, the joint prior distribution of $(Y_{c,0}, X_{c,0})$ is a truncated normal distribution with control parameters $\bm{\mu}_{early}$ and $\bm{\Sigma}_{early}$ given by
	\begin{align*}
    \left[ \begin{matrix}
	Y_{c,0} \\
	X_{c,0}
	\end{matrix} \right] \sim& \, TN\left( \bm{\mu}_{early} , 
	\bm{\Sigma}_{early} \right).
	\end{align*}
The truncation is such that $X_{c,0} \in [X_{0,low}, X_{0,up}]$ and $Y_{c,0} \in [Y_{0,low}, Y_{0,up}]$. 
These boundaries should be specified based on substantive knowledge about the ranges of $\mathbf{X}$ and $\mathbf{Y}$ and can vary with $c$, but this dependency is suppressed in the notation. 
The control parameters $\bm{\mu}_{early}$ and $\bm{\Sigma}_{early}$ are shared across all $c$.

Priors were chosen to be conjugate and diffuse for most parameters.
An informative prior was used for the autoregressive parameter $\rho$ in the imputation model for $\mathbf{Y} | \mathbf{X}$ to reflect the prior belief that $\mathbf{Y}$ is generally increasing over time for the motivating school enrollment data, but should generally be specified based on the data being imputed.
As the spline term $\beta f(X_{c,t})$ and the AR(1) term $\rho Y_{c,t-1}$ are both on the scale of $Y_{c,t}$, the parameters $\beta$ and $\rho$ can be interpreted as weights that represent the relative importance of the spline term and the AR(1) term.
The prior distribution for $\beta$ was chosen to be on the same order of magnitude as $\rho$ to reflect this interpretation.

The values for all hyperparameters should ideally be determined using prior expert knowledge.
In the absence of sufficient expertise to specify the prior distributions directly, researchers creating multiple imputations are still likely to have  non-expert knowledge of the substantive application that can be used to determine the appropriate order of magnitude for the parameters. 
Using the school enrollment data as an example, researchers may not have the expertise to specify the prior distributions for $\mu_{drift}$ and $\sigma^2_{drift}$ directly.
However, they may still have vague prior knowledge that the country-specific drift terms for GER should generally be positive and that increases in school enrollment happen slowly over time for most countries.

As an approximate quantification of this type of vague prior knowledge, we propose an algorithm for specifying diffuse priors dictated by data-based control parameters in the absence of prior expert knowledge. 
The data-based algorithm enables semi-automated specification of prior distributions using sample statistics calculated from the observed data.
The control parameters for the prior distributions of the $\mathbf{X}$ model ($\delta_X, \nu_{drift}, \zeta_{drift}, \text{and } \delta_{drift}$) are estimated using summary statistics for the observed first differences $X_{c,t} - X_{c,t-1}$ in $\mathbf{X}_{obs}$.
Similarly, the control parameters for the prior distributions of the $\mathbf{Y}|\mathbf{X}$ model ($ \delta_Y, \zeta_0, \text{and } \delta_0$) are estimated using summary statistics for the observed first differences $Y_{c,t} - Y_{c,t-1}$ in $\mathbf{Y}_{obs}$. 
The control parameters $\bm{\mu}_{early}$ and $\bm{\Sigma}_{early}$ for the prior distributions of $(Y_{c,0}, X_{c,0})$ are estimated using summary statistics for $\mathbf{Y}$ and $\mathbf{X}$ from a subset of the data that corresponds to ``early'' values of $t$, where the range of $t$ that is considered ``early'' is determined using substantive knowledge of the application.
Further details of the algorithm can be found in Section 1 of the Supplementary Material (\cite{LiuRaftery2025}), and an example illustrating how the algorithm is used to specify prior distributions for the school enrollment data can be found in Section \ref{sec:enroll_full}.
We note that use of a data-based algorithm to specify prior distributions results in an approximate posterior distribution rather than a fully Bayesian posterior distribution \citep{Darnieder2011}.

Although the proposed data-based algorithm is designed to specify prior distributions that are sufficiently diffuse such that the control parameters do not overwhelm the posterior inference following the philosophy of \cite{EdwardsLindmanSavage1963}, the prior distributions resulting from the algorithm should not be used blindly.
The prior distributions should be checked to ensure they are sensible for the substantive application and do not unnecessarily restrict the parameter space.
For the specification of $\bm{\mu}_{early}$ and $\bm{\Sigma}_{early}$, the data-based algorithm assumes that the time series for $\mathbf{X}$ and $\mathbf{Y}$ change smoothly over time and that the unobserved values $(Y_{c,0}, X_{c,0})$ are similar to the observed values in the early subset of data.
If there is reason to believe these assumptions are violated, the data-based algorithm should be used with caution.

\subsubsection{A-splines} \label{sec:aspline}
The nonlinear functions $f$ and $h$ in Equation \ref{eqn:Y_model} are estimated through spline regression using the complete cases in $(\mathbf{X}, \mathbf{Y})$.
To estimate $f$, the model $Y_{c,t} = f(X_{c,t}) + \varepsilon^f_{c,t}$ is fit with Gaussian errors using a B-spline of degree 1.
The residuals from the estimation of $f$ are then used to estimate $h$ by fitting the model $|Y_{c,t} - f(X_{c,t})| = h(X_{c,t}) + \varepsilon^h_{c,t}$ with Gaussian errors and using a B-spline of degree 1.
After estimation, $f$ is truncated to have range $[Y_{low}, Y_{up}]$ and $h$ is truncated to have range $[\epsilon, \infty)$, where $\epsilon$ is a small positive value.
The number and placement of knots for the B-splines used for $f$ and $h$ are selected using a method called adaptive splines, or A-splines (\cite{Goepp2018}).
A-splines automates the selection of knots using an iterative penalized likelihood approach and is implemented in the R package ``aspline'' (\cite{aspline}).

\subsection{Estimation}
\subsubsection{Model Estimation}
The MINTS model is estimated using a Markov chain Monte Carlo (MCMC) algorithm with Gibbs sampling and Metropolis-Hastings steps in R. 
Multiple imputations are created in two phases.
The parameters of the imputation model are first estimated in the estimation phase.
In the imputation phase, additional iterations from the same MCMC algorithm are run and used to create multiply imputed data sets.

Estimation of the imputation model parameters occurs simultaneously with estimation of the missing values in a similar fashion to the data augmentation algorithm of \citet{TannerWong1987}, with the MCMC algorithm resulting in samples from the joint posterior distribution of the imputation model parameters and the missing values given the observed data.
At each iteration of the MCMC algorithm, two steps are iterated until convergence is reached. 
First, values of the imputation model parameters are drawn from their posterior distributions given the observed data and the most recent estimates of the missing values.
Second, estimates for the missing values are drawn from their posterior distributions given the observed data and the previously drawn imputation model parameters.

The nonlinear functions $f$ and $h$ are estimated before the start of the MCMC algorithm using the complete cases in the data, pooled across countries and times. 
The same estimates of $f$ and $h$ are used throughout the estimation and imputation phases.
$f$ and $h$ are treated as deterministic within the MCMC algorithm, and uncertainty in the estimation of the nonlinear functions is not accounted for. 

Let $\bm{\theta}_X = (\bm{\gamma}, \sigma_X^2, \mathbf{X}_0, \mu_{drift}, \sigma^2_{drift})$ and $\bm{\theta}_Y = (\bm{\alpha}, \beta, \rho, \sigma_{Y}^2, \mathbf{Y}_0, \mu_0, \sigma^2_0)$ denote the parameters of the models for $\mathbf{X}$ and $\mathbf{Y} | \mathbf{X}$, respectively. 
The general approach of the MCMC algorithm proceeds as follows for iterations $i = 1, ..., n_{iter}$:
\begin{enumerate}
\item Draw $\bm{\theta}_X^{(i)}$ from $p(\bm{\theta}_X | \mathbf{X}_{obs}, \mathbf{X}_{mis}^{(i-1)})$
\item Let $j = 1, ..., J$ index the $X_{c,t}$ in $\mathbf{X}_{mis}$. 
For each $j$, draw $X^{(i)}_{j}$ from 
\begin{align*}
 p(X_{j} | \mathbf{X}_{obs}, X_1^{(i)}, X_2^{(i)}, ..., X_{j-1}^{(i)}, X_{j+1}^{(i-1)}, ..., X_J^{(i-1)}, \bm{\theta}_X^{(i)})
\end{align*}
\item Draw $\bm{\theta}_Y^{(i)}$ from $p(\bm{\theta}_Y | \mathbf{Y}_{obs}, \mathbf{Y}_{mis}^{(i-1)}, \mathbf{X}_{obs}, \mathbf{X}_{mis}^{(i)})$
\item Let $k = 1, ..., K$ index the $Y_{c,t}$ in $\mathbf{Y}_{mis}$. 
For each $k$, draw $Y^{(i)}_{k}$ from 
\begin{align*}
 p(Y_{k} | \mathbf{Y}_{obs}, Y_1^{(i)}, Y_2^{(i)}, ..., Y_{k-1}^{(i)}, Y_{k+1}^{(i-1)}, ..., Y_K^{(i-1)}, \mathbf{X}_{obs}, \mathbf{X}_{mis}^{(i)}, \bm{\theta}_Y^{(i)})
\end{align*} 
\end{enumerate}
The total number of iterations $n_{iter}$ used in the estimation phase is determined based on convergence diagnostics such as inspection of trace plots and evaluation of the diagnostics of \cite{RafteryLewis1996} and \cite{GelmanRubin1992}.
Complete details of the MCMC algorithm can be found in Section 2 of the Supplementary Material (\cite{LiuRaftery2025}).

\subsubsection{Imputation Procedure} \label{sec:imputation_procedure}
After the MCMC algorithm has converged for estimation of the imputation model parameters, the imputation phase begins.
All iterations of the MCMC that were required for convergence are treated as burn-in during the imputation phase. 
Imputed values for $\mathbf{X}_{mis}$ and $\mathbf{Y}_{mis}$ are created by continuing the MCMC algorithm with additional thinning steps.
Thinning of the MCMC chains is required during the imputation phase to ensure that the imputed data sets are approximately independent draws from the posterior predictive distribution of the missing data given the observed data under the model described in Section \ref{sec:MINTS_model} and priors described in Section \ref{sec:priors}.
The amount of thinning is chosen so that the autocorrelation of the imputed values is approximately zero. 

The number of iterations used in the imputation phase depends on the desired number of multiply imputed data sets, the number of chains of the MCMC, and the number of iterations used for thinning. 
To obtain $M$ multiply imputed data sets from $C$ chains with $n_{thin}$ iterations between imputed values, the MCMC algorithm is run for an additional $\frac{M}{C} \times n_{thin}$ iterations for each chain.

We note that although missing data is imputed for both $\mathbf{X}_{mis}$ and $\mathbf{Y}_{mis}$, the goal of MINTS is to create multiple imputations of $\mathbf{Y}$.
The auxiliary variable $\mathbf{X}$ is imputed out of necessity since $\mathbf{X}$ can also have missing data, but the imputed values of $\mathbf{X}$ are not intended to be used in subsequent analyses.

\section{Simulation Study} \label{sec:simulation}
We conducted a simulation study to evaluate how well the MINTS method performs for estimation of analysis models that are uncongenial to the imputation model, where uncongeniality arises because imputation of missing data and analysis of imputed data are conducted independently and information is not shared between imputation and analysis tasks. 
We refer to this validation exercise as ``analysis model validation.''

Analysis model validation was conducted for a simulated data set where a nonlinear relationship was simulated between variables. 
We considered nine experiments corresponding to three rates of simulated missingness and three missing data mechanisms.
Each experiment was replicated $N_{rep} = 1000$ times.
The average performance across replications for MINTS was compared with the performance of existing multiple imputation methods for hierarchical time series data. 
Details of the simulation study using the nonlinear simulated data are presented in this section, while details of an analogous simulation study using linear simulated data are available in Sections 5 and 6 of the Supplementary Material (\cite{LiuRaftery2025}). 

\subsection{Data Generation}
Variables $\mathbf{X}$ and $\mathbf{Y}$ were generated for 20 countries and 30 years for a total sample size of 600 country-years. 
$\mathbf{Y}$ is the variable of interest for substantive analyses, while $\mathbf{X}$ is an auxiliary variable that is only of interest for imputation of $\mathbf{Y}$. 

$\mathbf{X}$ was simulated independently for each country and is bounded in $[0, 100]$. For country $c$, 
\begin{align*}
X_{c,1} \sim& \, U(X_{1,low}, X_{1,up}), \\
X_{c, t+1} \sim& \, TN_{[0, 100]}(X_{c, t} + \gamma_c, \sigma^2) \text{  for $t = 1, ..., 29$},
\end{align*}
where $TN_{[0, 100]}$ refers to the truncated normal distribution with support $[0, 100]$ and $\gamma_c$ is a country-specific drift term. 
For the nonlinear simulated data, we set
$X_{1,low} = 0$, 
$X_{1,up} = 25$, 
$\sigma^2 = 1$, 
and $\gamma_c \sim U(1, 3)$. 

$\mathbf{Y}$ was simulated to have a nonlinear relationship with $\mathbf{X}$ and is bounded as $Y_{c,t} \in [0, min(X_{c,t}, 60)]$. 
For country $c$,
\begin{align*}
Y_{c,t} \sim& \, TN_{[0, \text{min}(X_{c,t}, 60)]} \Bigg(\alpha_c +  \frac{40}{\big(1 + \exp\big(- \frac{X_{c,t} - 60}{8} \big) \big)} + 3 \log(X_{c,t}), 1^2 \Bigg), \\
\alpha_c \sim& \, U(0, 5),
\end{align*}
where $TN_{[0, \text{min}(X_{c,t}, 60)]}$ refers to the truncated normal distribution with support $[0, \text{min}(X_{c,t}, 60)]$.
$\mathbf{X}$ and $\mathbf{Y}$ were constructed to have a generally monotonically increasing relationship similar to the relationship observed for the enrollment data. 
We assume that the bounds of $\mathbf{X}$ and $\mathbf{Y}$ are known to the imputer.

\subsection{Analysis Model}
We focused on the setting where the analysis model is uncongenial to the imputation model. 
The variable $\mathbf{Z}$ is treated as the outcome variable and was simulated to have a linear relationship with $\mathbf{Y}$.
For each country $c$ and year $t$, 
\begin{align*}
Z_{c,t} \sim& \, N(\eta_c + 2 Y_{c,t}, 10^2), \\
\eta_c \sim& \, U(0, 15).
\end{align*}
The analysis model is the linear regression of $\mathbf{Z}$ on $\mathbf{Y}$.
The parameter of interest is $\omega_1$, the coefficient on $\mathbf{Y}$ in the regression
\begin{align*}
Z_{c,t} =& \, \omega_0 + \omega_1 Y_{c,t} + \varepsilon_{c,t}^{\omega}, \\
\varepsilon_{c,t}^{\omega} \sim& N(0, \sigma_{\varepsilon_{\omega}}^2).
\end{align*}

This analysis model is uncongenial to the imputation model due to the presence of $\mathbf{Z}$, which is not included in any of the imputation models considered. 
We assume $\mathbf{X}$ is only available during the imputation task and is not possible to include in subsequent analyses. 
For example, this could occur if the refined measure $\mathbf{Y}$ comes from survey data and the coarse measure $\mathbf{X}$ comes from confidential administrative records data.
Multiply imputed values of $\mathbf{Y}$ are desired for public release, but $\mathbf{X}$ cannot be made publicly available. 
We also assume $\mathbf{Z}$ is only available during the analysis task and is not possible to include in the imputation model. 
This could occur if $\mathbf{Z}$ is collected by the analyst after the multiple imputation task is complete or because $\mathbf{Z}$ is unavailable to the imputer for data confidentiality reasons.
We note that omitted variable bias is likely to occur in this setting due to the omission of $\mathbf{Z}$ in the imputation model, so a data augmentation approach based on MINTS that incorporates $\mathbf{Z}$ into the imputation model is also considered in the validation exercises for comparison purposes. 

Additional validation results for a random intercept analysis model can be found in Section 4 of the Supplementary Material (\cite{LiuRaftery2025}).

\subsection{Analysis Model Validation Procedure} \label{sec:mechanisms}
Data was simulated as missing following the three missing data mechanisms of \cite{Rubin1976}: Missing Completely At Random (MCAR), Missing At Random (MAR), and Missing Not at Random (MNAR).
For a variable $\mathbf{X}$, MCAR occurs when $P(\mathbf{R}^X | \mathbf{X}) = P(\mathbf{R}^X)$. 
Data was simulated as missing under MCAR by assuming that each observation has the same probability of being missing. 
MAR occurs when $P(\mathbf{R}^X | \mathbf{X}) = P(\mathbf{R}^X | \mathbf{X}_{obs})$. 
Data was simulated as missing under MAR by assuming that observations in earlier years are more likely to be missing. 
Finally, MNAR occurs when the probability of being missing depends on both the missing and the observed data and $P(\mathbf{R}^X | \mathbf{X})$ cannot be simplified further.
Data was simulated as missing under MNAR by assuming that the probability of $\mathbf{X}$ being missing depends on the value of $\mathbf{X}$.
Details of the MAR and MNAR implementations can be found in Section 3 of the Supplementary Material (\cite{LiuRaftery2025}). 

For each missing data mechanism, data was simulated as missing at the 10\%, 40\%, and 80\% rates. 
Each combination of missing data mechanism and rate was implemented simultaneously for $\mathbf{X}$ and $\mathbf{Y}$ and defines an experiment.
For example, the MCAR 10\% experiment corresponds to the setting where 10\% of $\mathbf{X}$ was simulated as missing under MCAR and 10\% of $\mathbf{Y}$ was simulated as missing under MCAR. 
We note that while 80\% is a high rate of missingness, it is similar to the observed rate of 73.0\% missingness for NER in the school enrollment data set. 

All multiple imputation methods compared in the validation exercises assume that the missing data mechanism is MAR and that there is sufficient information from the observed data to adequately estimate the relationship between $\mathbf{X}$ and $\mathbf{Y}$. 
The experiments where data is simulated under MNAR act as a sensitivity analysis to evaluate each method's performance when the MAR assumption is violated.
Similarly, the experiments using the 80\% rate of simulated missingness are considered to assess how well each method performs when there is a very large amount of missing data. 
In practice, substantive knowledge about the variables with missing data should be used to determine if multiple imputation is appropriate for use if the missing data mechanism is suspected to be MNAR or if there are large amounts of missing data.

For each experiment, the analysis model validation procedure is
\begin{enumerate}
\item  For replication $r = 1, ..., N_{rep}$,
	\begin{enumerate}
	\item Simulate missing values according to the experiment's missing data mechanism and rate to separate the data into ``observed'' and ``missing'' data
	\item Run the multiple imputation procedure using the observed data to create $M = 40$ completed data sets. Completed data sets consist of the observed data and the imputed values for the missing data.
	\item Estimate quantities of interest $Q$ using each of the $M$ completed data sets
	\item Pool the estimates of $Q$ and $SE(Q)$ across the $M$ completed data sets using combining rules from Rubin (1987) to obtain the pooled estimates $\bar{Q}_r$ and $SE(\bar{Q}_r)$
	\end{enumerate}
\item Calculate the true value of $Q$ using the full data set 
\item Calculate evaluation metrics for the pooled estimates $\bar{Q}_r$ averaged across replications
\end{enumerate}

The number of imputations $M = 40$ was chosen to balance between having a large enough number of imputations to guarantee minimal contribution of simulation error to the variability of estimands and having a small enough number of imputations to be computationally feasible.

Pooled estimates for each scalar quantity of interest $Q$ are created using combining rules from Rubin (1987) in Step 1(d) of the validation procedure.
Let $\hat{Q}_m$ denote the point estimate of $Q$ from imputation $m$ and let $\hat{U}_m$ denote its associated variance. 
The pooled point estimate of $Q$ across all $M$ imputations is $\bar{Q}_M = \frac{1}{M} \sum_{m=1}^M \hat{Q}_m$.
The pooled variance estimate of $\bar{Q}_M$ is $T_M = \bar{U}_M + \left(1 + \frac{1}{M} \right) B_M$, where $\bar{U}_M = \frac{1}{M} \sum_{m=1}^M \hat{U}_m$ is the average within-imputation variance and $B_M = \frac{1}{M-1} \sum_{m=1}^M ( \hat{Q}_m - \bar{Q}_M )^T ( \hat{Q}_m - \bar{Q}_M )$ is the between-imputation variance.
The 95\% confidence interval for the pooled estimate $\bar{Q}_M$ is constructed using $\bar{Q}_M \pm t_{\nu}^* \sqrt{T_M}$, where $t_{\nu}^*$ is the critical value for 95\% confidence from the $t$ distribution with $\nu$ degrees of freedom.
The degrees of freedom $\nu$ for finite number of imputations $M$ is 
$\nu = (M - 1) (1 + \frac{1}{r_M})^2$, where $r_M$ is the relative increase in variance due to nonresponse given by $r_M = \left(1 + \frac{1}{M} \right) \frac{B_M}{\bar{U}_M}$.

The performance of the point estimates $\bar{Q}_M$ was evaluated using the mean absolute error (MAE), calculated as $\frac{1}{N_{rep}} \sum_r |\bar{Q}_M - Q|$ where the sum is taken over all replications within each experiment. 
The performance of the 95\% confidence intervals for the pooled estimates $\bar{Q}_M$ was evaluated using the mean coverage across all replications within each experiment, calculated as the proportion of intervals that contained the true value.
We also evaluated the mean fraction of Fisher information about $Q$ that is missing due to nonresponse, which we abbreviate as FMI for Fraction of Missing Information.
In each replication of each experiment, FMI is estimated as
\begin{align*}
\text{FMI} =& \frac{r_M + \frac{2}{\nu + 3}}{r_M + 1}.
\end{align*}
The mean FMI across all replications within each experiment enables assessment of the amount of information about $Q$ that is lost due to the presence of missing data (\cite{Schafer1997, SavaleiRhemtulla2012}). 
We note that with $M = 40$ imputations, the estimates of FMI may be noisy, so in this case FMI  should only be interpreted as an exploratory diagnostic  (\cite{Bodner2008, Enders2010}).

\subsection{Model Implementation} \label{sec:models}
For each replication of each experiment, we created 40 imputations using the MINTS method by running 10 chains of the MCMC algorithm.  
The bounds of the model for $\mathbf{X}$ were set as $X_{low} = 0$ and $X_{up} = 100$, and the bounds of the model for $\mathbf{Y} | \mathbf{X}$ were set as $Y_{low} = 0$ and $Y_{up} = \text{min}(X_{c,t}, 60)$.
During the estimation phase, the MCMC algorithm was run for enough iterations to ensure convergence of all imputation model parameters. 
The number of iterations differed across experiments, but ranged from 10,000 to 25,000 iterations per chain. 
During the imputation phase, an additional 4,000 iterations was run for each chain and four iterations from each chain were selected as the imputed values. 
The iterations selected as the final imputed values were chosen to be 1,000 iterations apart to ensure autocorrelation was close to zero following the procedure described in Section \ref{sec:imputation_procedure}. 

We compared the MINTS method to six models based on existing methods for multiple imputation.
As the existing methods assume a linear relationship between variables with missing data, the simulated data was transformed prior to creating multiple imputations using a Box-Cox transformation.
For each of the existing methods, $\mathbf{Y}$ was first transformed using the cube root as $\mathbf{Y}^* = \mathbf{Y}^{1/3}$.
Imputations were created on the transformed scale, and the resulting multiply imputed values of $\mathbf{Y}^*$ were transformed back to the original scale of $\mathbf{Y}$. 

Three of the existing multiple imputation models considered are based on the MICE methodology as implemented in the R package ``mice'' (\cite{vanBuurenOudshoorn2011}).\footnote{mice version 3.14.0 used}
MICE uses the FCS algorithm, which allows for the specification of separate univariate conditional models for each variable with missing data. 
The univariate conditional models include all available variables as predictors, for example, the model for $\mathbf{X}$ includes $\mathbf{Y}$, year, and country as predictors.  
We first considered the default imputation method for continuous data in the mice package, which is predictive mean matching.
We refer to this model as MICE PMM.
Unlike the other approaches considered, MICE PMM does not explicitly account for the hierarchical structure of the data but is included as a baseline for comparison.

We evaluated two models within the MICE framework that account for the hierarchical structure of the data using the method of \cite{Schafer1997TechnicalReport} and \cite{SchaferYucel2002}, referred to as the pan method following its implementation in the R package ``pan'' (\cite{pan}).
The function \textit{mice.impute.2l.pan} within the mice package uses a Gibbs sampler to estimate the conditional linear mixed effects model with homogeneous within-group variances for each variable with missing data given the other variables.
We evaluated a model that includes a country-specific intercept and fixed effects for all covariates in the imputation model.
We refer to this model as pan Fixed Effects.
For example, imputed values for $X_{c,t} | Y_{c,t}$ are modeled with a country-specific intercept, a fixed effect of year, a fixed effect of $Y_{c,t}$, and a homogeneous normally distributed error term. 
We also considered the pan Random Effects model, which adds random effects of all covariates to the model used in pan Fixed Effects. 

We evaluated three models using the Amelia II methodology as implemented in the R package ``Amelia'' (\cite{HonakerKing2010}).\footnote{Amelia version 1.80 used}
For complete data $(\mathbf{X}, \mathbf{Y})$, Amelia assumes the joint distribution is multivariate normal. 
The parameters of the joint distribution are estimated using a combination of bootstrapping and an EM algorithm. 
Amelia is designed specifically for imputation of hierarchical time series data, which Honaker and King refer to as time series cross-sectional data, through modeling features such as smooth trends over time and allowing for country-specific effects.
For our comparisons, we evaluated three implementations of Amelia that were chosen to use the simplest form of the time-series-cross-sectional modeling features in Amelia: a time-series (TS) model, a cross-sectional (CS) model, and a time-series-cross-sectional (TSCS) model.
All three of the imputation models are constructed by adding terms to the default multivariate normal joint model for $(\mathbf{X}, \mathbf{Y})$.
In the Amelia TS model, a linear effect of time is added.
In the Amelia CS model, a country-specific intercept term is added.
In the Amelia TSCS model, a country-specific intercept term and a country-specific linear effect of time are added.

Imputations were created using the default settings in mice and Amelia, with the exception of setting the number of imputations as $M = 40$, specifying the form of the imputation models as described above, and specifying bounds of $\mathbf{X} \in [0, 100]$ and $\mathbf{Y} \in [0, 60]$. 
Scalar bounds were used for $\mathbf{Y}$ as the mice and Amelia packages do not allow for variable bounds.

We also considered a data augmentation approach that extends the MINTS method to enable estimation of the parameters of the analysis model through the decomposition
\begin{align*}
p(\mathbf{Z}, \mathbf{Y}, \mathbf{X} | \bm{\theta}_Z, \bm{\theta}_Y, \bm{\theta}_X) =& p(\mathbf{Z}| \mathbf{Y}, \mathbf{X}, \bm{\theta}_Z) p(\mathbf{Y} | \mathbf{X}, \bm{\theta}_Y) p(\mathbf{X} | \bm{\theta}_X),
\end{align*}
where $\bm{\theta}_Z$ denotes the parameters of the analysis model.
The data augmentation approach uses the same models for $\mathbf{Y} | \mathbf{X}$ and $\mathbf{X}$ as MINTS, while the model for $\mathbf{Z}| \mathbf{Y}, \mathbf{X}$ is given by the analysis model used in the validation exercise. 
We abbreviate the data augmentation approach based on MINTS as the MINTS DA method.
MINTS DA can be used to directly sample from the posterior distributions for the parameters of the analysis model rather than needing to rely on the multiple imputation framework.
Medians of the posterior distributions for parameters of interest were used for point estimates, while the 0.025th and 0.975th quantiles of the posterior distributions were used for interval estimates. 
As the estimates of the parameters of interest are not constructed using Rubin's pooling rules, we did not calculate estimates of the Fraction of Missing Information.
We note that MINTS DA is considered only for comparison purposes as a method that directly incorporates information about $\mathbf{Z}$ when handling missing data in $\mathbf{Y}$ and $\mathbf{X}$.
In practice, the data augmentation approach is not possible to implement in our motivating setting where the substantive analysis of interest and imputation of missing data are conducted independently, and information about $\mathbf{Z}$ and $\mathbf{X}$ are not shared between analyst and imputer.

\subsection{Analysis Model Validation Results}
For the linear regression analysis model validation, we evaluated how well each multiple imputation method performs for estimation of $Q = \omega_1$, the regression coefficient on $\mathbf{Y}$ in the linear regression of $\mathbf{Z}$ on $\mathbf{Y}$. 
Table \ref{tab:val3_lm} summarizes the results of this validation exercise.

Overall, we found MINTS results in the best balance between MAE, coverage, and FMI.
Out of the multiple imputation methods considered, MINTS has the smallest MAE in all experiments except MAR 10\% and MAR 40\%, where MINTS has the second smallest MAE behind MICE PMM. 
MINTS has reasonably close to nominal coverage for 95\% intervals at the 10\% and 40\% rates of simulated missingness, but suffers from undercoverage at the 80\% rate.
The Amelia TSCS method has closer to nominal coverage than MINTS in the MCAR 80\% and MNAR 80\% experiments at the expense of higher MAE, but MINTS has the closest to nominal coverage in the MAR 80\% experiment. 

MINTS has the smallest FMI in eight out of nine experiments.
As expected, FMI is much larger for all methods at the 80\% rate, with FMI $> 50\%$ for several methods.
This is higher than is typical for the setting of sample survey data that multiple imputation was originally designed for, where FMI is usually $\leq 30\%$ (Rubin 2007).
Given the high FMI, it is thus unsurprising how poorly all multiple imputation methods perform in the experiments at the 80\% rate.
In particular, the spline estimation used in MINTS suffers from overfitting at the 80\% rate due to the small number of complete cases available to estimate the splines $f$ and $h$, with some replications having less than 15 complete cases. 

None of the existing multiple imputation methods perform consistently well across experiments for estimation of $\omega_1$.
While MICE PMM can outperform the methods explicitly designed for hierarchical time series data at the lower rates of missingness, the performance of MICE PMM suffers greatly at the highest rate of missingness.
The pan and Amelia methods perform similarly to one another at the 10\% rate, but all have larger MAE than MICE PMM. 
Performance of the pan and Amelia methods generally is best under MCAR, with pan Random Effects and Amelia TSCS performing the best of the group in terms of MAE and coverage. 

Multiple imputation using the MINTS method generally has better performance than the data augmentation approach based on MINTS at the lower rates of simulated missingness. 
However, the data augmentation approach has smaller MAE than all multiple imputation methods at the highest rate of simulated missingness. 
The multiple imputation methods all have substantial undercoverage in the MAR 80\% and MNAR 80\% experiments, while the data augmentation approach maintains close to nominal coverage.  
These findings suggest that omission of $\mathbf{Z}$ from the imputation model is likely a major source of bias and undercoverage for the multiple imputation methods when there is a small number of observed values.

\begin{table}[!htb]
\caption{Summary of analysis model validation for nonlinear simulated data for $Q = \omega_1$, the regression coefficient on $Y$ in the linear regression of $Z$ on $Y$. MAE denotes mean absolute error, Cvg denotes the average coverage of 95\% intervals as a percentage, and FMI denotes the fraction of missing information as a percentage. MAE is multiplied by 100 before reporting. Results are averaged over the 1000 replications of each experiment. The true value of $Q$ is 2.060.} \label{tab:val3_lm}
\centerline{
\small
\begin{tabular}{cc|rrr|rrr|rrr}
  \hline
Simulated  & \multirow{2}{*}{Method}  & \multicolumn{3}{c|}{\textbf{MCAR}}  & \multicolumn{3}{c|}{\textbf{MAR}} & \multicolumn{3}{c}{\textbf{MNAR}} \\
 \multicolumn{1}{c}{Missingness Rate} &  & MAE & Cvg & FMI & MAE & Cvg & FMI & MAE & Cvg & FMI \\
   \hline 
\multirow{8}{*}{\textbf{10\%}}  &  MICE PMM & 0.85 & 100.0 & 5.1 & \textbf{0.38} & 100.0 & 1.6 & 0.49 & 100.0 & 2.6 \\ 
&  pan Fixed & 1.36 & 100.0 & 8.6 & 1.69 & 100.0 & 1.9 & 1.06 & 100.0 & 3.2 \\ 
&  pan Random & 1.08 & 100.0 & 7.6 & 1.28 & 100.0 & 1.4 & 0.77 & 100.0 & 2.3 \\ 
&  Amelia TS & 1.60 & 100.0 & 10.9 & 1.89 & 100.0 & 2.6 & 1.13 & 100.0 & 3.8 \\ 
&  Amelia CS & 1.38 & 100.0 & 7.9 & 2.26 & 100.0 & 4.4 & 1.29 & 100.0 & 3.7 \\ 
&  Amelia TSCS & 1.28 & 100.0 & 8.1 & 1.40 & 100.0 & 1.3 & 0.70 & 100.0 & 2.1 \\ 
&  MINTS & \textbf{0.29} & 100.0 & \textbf{0.5} & 0.56 & 100.0 & \textbf{0.5} & \textbf{0.29} & 100.0 & \textbf{0.4} \\ 
& MINTS DA & 1.00 & 100.0 &  & 0.67 & 100.0 &  & 0.97 & 100.0 &  \\  
 & &  & & &  &  &  &  &  &  \\ 
\multirow{8}{*}{\textbf{40\%}} & MICE PMM & 12.60 & 40.0 & 36.6 & 2.22 & 100.0 & 13.9 & 8.45 & 77.8 & 28.4 \\ 
&  pan Fixed & 6.87 & 91.2 & 33.2 & 10.13 & 15.3 & 14.9 & 8.57 & 66.7 & 22.6 \\ 
&  pan Random & 5.33 & \textbf{94.3} & 36.9 & 6.54 & 88.8 & 12.1 & 4.19 & 99.9 & 15.9 \\ 
&  Amelia TS & 7.18 & \textbf{94.3} & 29.0 & 10.63 & 13.9 & 17.1 & 7.83 & 82.4 & 19.9 \\ 
&  Amelia CS & 14.25 & 35.2 & 26.8 & 12.11 & 33.4 & 19.0 & 12.17 & 21.9 & 21.1 \\ 
&  Amelia TSCS & 4.70 & 93.6 & 38.3 & 6.96 & 81.2 & 17.1 & 3.50 & 99.4 & 17.2 \\ 
&  MINTS  & \textbf{1.21} & 100.0 & \textbf{3.9} & 3.02 & 99.9 & \textbf{4.4} & \textbf{1.49} & 100.0 & \textbf{2.8} \\ 
& MINTS DA &  2.38 & 91.3 &  & \textbf{1.80} & \textbf{98.7} &  & 2.21 & \textbf{98.5} &  \\ 
 & &  & &  &  &  &  &  &  &  \\ 
\multirow{7}{*}{\textbf{80\%}} & MICE PMM & 58.07 & 0.1 & 60.1 & 24.73 & 14.1 & 52.7 & 56.68 & 0.0 & 53.9 \\ 
&   pan Fixed & 32.27 & 8.6 & 65.2 & 29.84 & 0.0 & 47.7 & 37.56 & 0.0 & 58.8 \\ 
&   pan Random & 42.19 & 34.6 & 87.5 & 38.34 & 17.5 & 84.3 & 45.67 & 47.5 & 86.1 \\ 
&   Amelia TS & 20.38 & 69.3 & 54.4 & 33.08 & 0.0 & \textbf{42.3} & 24.87 & 3.7 & 39.5 \\ 
&   Amelia CS & 91.98 & 0.0 & 72.7 & 70.19 & 0.0 & 79.0 & 63.14 & 0.0 & 64.5 \\ 
&   Amelia TSCS & 30.58 & \textbf{91.0} & 93.8 & 40.11 & 25.2 & 89.1 & 27.10 & 75.4 & 89.1 \\ 
&  MINTS  & 7.62 & 81.0 & \textbf{38.7} & 13.14 & 47.5 & 47.0 & 9.71 & 62.6 & \textbf{35.6} \\ 
& MINTS DA & \textbf{4.05} & 88.8 &  & \textbf{4.53} & \textbf{99.4} &  & \textbf{3.36} & \textbf{96.9} &  \\ 
   \hline
\end{tabular}
}
\end{table}

We also conducted a simulation study for data simulated to follow a linear relationship.
For the linear simulated data, we considered two uncongenial and two congenial analysis models.
Although our primary focus is on the uncongenial setting, the analysis model validation using congenial analysis models and linear simulated data allows us to compare the performance of MINTS with the existing imputation methods in an ``ideal'' setting for multiple imputation.
We found that the existing imputation methods perform better in the analysis model validation for the linear simulated data compared to the nonlinear simulated data. 
However, for estimation of uncongenial analysis models, we found MINTS still outperforms the existing methods in the linear setting at the 10\% and 40\% rates of simulated missingness. 
All imputation methods were found to perform well for estimation of congenial analysis models using linear simulated data, with no method consistently standing out from the others across experiments. 
Details of the simulation study using linear simulated data can be found in Sections 5 and 6 of the Supplementary Material (\cite{LiuRaftery2025}).

\section{Application to Enrollment Data} \label{sec:enroll}
We further evaluated the performance of MINTS by revisiting the secondary school enrollment data that was described in Section \ref{sec:data}.
We conducted two validation exercises using the school enrollment data by simulating additional country-years as missing.
In the first validation exercise, we evaluated how well MINTS performs for estimation of parameters of interest $Q$ for uncongenial analysis models. 
This is analogous to the analysis model validation that was conducted for the simulation study. 

In the second validation exercise, we evaluated the predictive performance of MINTS for predicting left-out observations of NER. 
Out-of-sample validation for prediction of individual missing values is less frequently conducted for multiple imputation methods compared to analysis model validation, but has been considered by \cite{Gelman1998, HonakerKing2010, Nguyen2017}; among others. 
Although the primary goal of multiple imputation is to create valid estimates of parameters of interest in the presence of missing data rather than recovering the missing values (\cite{Rubin1996}), the prediction of individual missing values can still be of great interest in practice.
A multiple imputation method that can perform well for prediction of individual missing values and for creating multiply imputed estimates of quantities of interest is thus of increased utility. 
We refer to this second validation exercise as ``out-of-sample validation.''

Finally, we applied the MINTS method to the full school enrollment data set without simulating any additional missing values and created 40 multiple imputations for the country-years of NER that are missing in the original data set.
We also estimated a substantive analysis model of interest using the multiply imputed values of NER created using MINTS and compared these results with the estimates obtained using the other methods considered in the validation exercises. 

\subsection{Validation Procedure}
For both validation exercises using the school enrollment data, we considered eight experiments.
Parameters varied in both validation exercises were the rate of simulated missingness (10\%, 40\%, 80\%) and the missing data mechanism (MCAR, MAR, and MNAR), where the same missing data mechanisms used in the simulation study and described in Section 3 of the Supplementary Material (\cite{LiuRaftery2025}) were also used for the enrollment data.
We did not consider the MNAR 80\% experiment for the enrollment data as this resulted in the majority of countries having no observations for NER or GER. 

As we simulated additional missing values in a data set that began with missing values, the overall rate of missingness for each experiment is larger than the rate of simulated missingness. 
For example, the experiments with a 40\% rate of simulated missingness correspond to an overall rate of missingness of 84.8\% for NER.
Similarly, the simulated missing data mechanisms used in the validation exercises describe the missing data mechanism only for observations that are simulated as missing. 
The true missing data mechanism for the observations that began as missing in the original enrollment data set is unknown. 

For the analysis model validation, each experiment was replicated $N_{rep} = 100$ times following an analogous procedure as was described in Section \ref{sec:mechanisms} for the simulation study.
A major difference in the analysis model validation procedure for the enrollment data is the need to distinguish between the country-years that start out as missing and the country-years that are simulated as missing, where only the country-years that are simulated as missing are used for evaluation purposes.
To facilitate this distinction, in each replication of each experiment we
separated the enrollment data into ``started-as-missing'', ``observed'', and ``simulated-as-missing'' sets.
The observed set is the training set for model estimation, while the simulated-as-missing set is the testing set for validation. 
Imputed values were still created for the started-as-missing set, but we are unable to evaluate the performance of these imputed values as the true values for the started-as-missing set are unknown.

For the out-of-sample validation exercise, we considered one replication of each experiment. 
We used the same distinction between the country-years that started as missing and the country-years that were simulated as missing, where only the country-years that were simulated as missing are included in the testing set for validation.  
The out-of-sample validation exercise compares performance metrics for prediction of missing values for NER averaged over all country-years in the testing set. 

In both validation exercises, we compared the performance of MINTS with the  MICE PMM, pan Fixed Effects, pan Random Effects, Amelia TS, Amelia CS, and Amelia TSCS models that were described in Section \ref{sec:models}.
For the analysis model validation exercise, we additionally compared the multiple imputation methods with the data augmentation approach described in Section \ref{sec:models}.

\subsection{Analysis Model Validation}
\subsubsection{Analysis Model} \label{sec:enroll_analysis}
We are interested in estimating the relationship between NER and the Total Fertility Rate (TFR), which is a period measure of the expected number of children a woman would bear in her lifetime if she were to experience the period-specific fertility rates at each age and if she lived through the reproductive age range of 15--49. 
There is a well-established negative association between education and fertility in the high-fertility setting (\cite{Hirschman1994}).
One mechanism through which education is posited to have a negative effect on fertility is through the educational enrollment of children, where increased  enrollment increases the cost of raising children (\cite{AxinnBarber2001}).
Children who are enrolled in school have reduced capacity for work and may incur increased costs for caregivers through fees related to tuition, uniforms, and textbooks (\cite{EasterlinCrimmins1985, Caldwell1982, Caldwell1985}).
To estimate this relationship, we use annual estimates of TFR from the 2022 revision of the United Nations \textit{World Population Prospects}  \citepalias{WPP2022}.
We restrict our analyses to the high-fertility context, defined here as the years where a country has TFR $> 2.5$ children per woman. 

The analysis model is the linear regression of TFR on NER, where the quantity of interest $Q = \beta_1$ is the regression coefficient on NER in the regression
	\begin{align*}
	\text{TFR}_{c,t} =& \beta_0 + \beta_1 \text{NER}_{c,t} + \varepsilon_{c,t}^{\beta}, \\
	\varepsilon_{c,t}^{\beta} \sim& N(0, \sigma_{\varepsilon_{\beta}}^2).
	\end{align*}
For each replication of each experiment, the analysis model is estimated using only the country-years that were simulated as missing and where TFR $>2.5$.
We also conducted analysis model validation using a random intercept model for TFR on NER, results of which are available in Section 4 of the Supplementary Material (\cite{LiuRaftery2025}). 
	
In this validation exercise, the analysis model is uncongenial to all imputation models considered due to the omission of the analysis model outcome variable $\mathbf{Z} = \text{TFR}$ from the imputation models. 
This reflect our motivating setting where analysis of the multiply imputed values occurs independently from multiple imputation, and information about the outcome variable is not known during the imputation task.
For comparison purposes, we also consider a data augmentation algorithm based on MINTS to represent the setting where imputation of missing values and estimation of the substantive analysis model can be conducted simultaneously.

\subsubsection{Model Implementation}
GER is the auxiliary variable $\mathbf{X}$ and NER is the variable of interest $\mathbf{Y}$. 
To satisfy the assumption of the MINTS method that the auxiliary variable is observed at least once for each country, we excluded all countries that have no observed values for GER.
The bounds of the MINTS imputation model were set based on substantive knowledge of the possible ranges of NER and GER: NER is a percentage, GER is a positive rate, and NER$_{c,t} \leq$ GER$_{c,t}$. 
The bounds of the model for $\mathbf{X}$ are set as $X_{low} = 0$ and $X_{up} = \infty$, while the bounds of the model for $\mathbf{Y} | \mathbf{X}$ were set as $Y_{low} = 0$ and $Y_{up} = \text{min}(X_{c,t}, 100)$.
Imputations were created using MINTS for each replication of each experiment by running 10 chains of the MCMC algorithm until convergence was reached in the estimation phase. 
The total number of iterations differed across experiments, with the experiments with larger rates of simulated missingness requiring a larger number of iterations to achieve convergence. 
After burn-in, the number of iterations per chain ranged from 5,000 to 35,000.
In the imputation phase, an additional 4,000 iterations was run for each chain and four iterations were selected from each chain to produce a total of $M = 40$ completed data sets following the procedure described in Section \ref{sec:imputation_procedure}. 

Multiple imputations were created using the models based on the MICE and Amelia methods following the same implementation as described in Section \ref{sec:models} with two exceptions.
First, the school enrollment data was not transformed prior to using the MICE and Amelia methods, as we were unable to find a transformation that substantially improved linearity for the relationship between $\mathbf{X}$ and $\mathbf{Y}$.
Second, the bounds of each model were specified as $\mathbf{X} \in [0, \infty)$ and $\mathbf{Y} \in [0, 100]$.
The data augmentation method based on MINTS was also implemented as described in Section \ref{sec:models}.

\subsubsection{Results}
Table \ref{tab:val2_lm} summarizes the results of the analysis model validation for estimation of $Q = \beta_1$, the regression coefficient on NER in the linear regression of TFR on NER. 
MINTS has the smallest MAE out of the multiple imputation methods considered in all but the MAR 80\% experiment.
MINTS also has the smallest FMI in all experiments and close to nominal coverage in all but the MNAR 40\% experiment.
Out of the previously existing methods, pan Fixed Effects performs the best overall with the smallest MAE in the MAR 80\% experiment and the closest to nominal coverage in the MNAR 40\% experiment. 
The data augmentation approach based on MINTS has the smallest MAE and closest to nominal coverage in the MNAR experiments, but falls behind several multiple imputation methods in the experiments where the MAR assumption is satisfied.

\begin{table}[!htb]
\caption{Summary of analysis model validation for enrollment data for $Q = \beta_1$, the regression coefficient on NER in the linear regression of TFR on NER. MAE denotes mean absolute error, Cvg denotes the average coverage of 95\% intervals as a percentage, and FMI denotes the fraction of missing information as a percentage. MAE is multiplied by 100 before reporting. Results are averaged over the 100 replications of each experiment. The value of $Q$ estimated using the observed country-years from the full enrollment data set is -0.043.} \label{tab:val2_lm}
\centerline{
\small
\begin{tabular}{cc|rrr|rrr|rrr}
  \hline
Simulated  & \multirow{2}{*}{Method}  & \multicolumn{3}{c|}{\textbf{MCAR}}  & \multicolumn{3}{c|}{\textbf{MAR}} & \multicolumn{3}{c}{\textbf{MNAR}} \\
 \multicolumn{1}{c}{Missingness Rate} &  & MAE & Cvg & FMI & MAE & Cvg & FMI & MAE & Cvg & FMI \\
   \hline 
\multirow{8}{*}{\textbf{10\%}}  &  MICE PMM & 0.61 & 90 & 20.9 & 0.86 & 71 & 26.3 & 7.29 & 0 & 28.2 \\ 
&  pan Fixed & 0.08 & 100 & 7.2 & 0.11 & 100 & 6.8 & 2.69 & 0 & 29.9 \\ 
&  pan Random & 0.07 & 100 & 5.3 & 0.12 & 100 & 7.8 & 4.08 & 2 & 88.4 \\ 
&  Amelia TS & 0.29 & 100 & 24.2 & 0.51 & \textbf{99} & 26.7 & 7.16 & 0 & 30.6 \\ 
&  Amelia CS & 0.21 & 100 & 11.4 & 0.40 & \textbf{99} & 11.9 & 6.03 & 0 & 58.9 \\ 
&  Amelia TSCS & 0.09 & 100 & 8.3 & 0.16 & 100 & 12.9 & 6.42 & 0 & 69.2 \\ 
&  MINTS & \textbf{0.04} & 100 & \textbf{2.2} & \textbf{0.08} & 100 & \textbf{2.4} & 0.42 & \textbf{100} & \textbf{10.8} \\ 
& MINTS DA & 0.28 & 100 &  & 0.23 & 100 &  & \textbf{0.34} & \textbf{100} &  \\ 
 & & & &  &  &  &  &  &  &  \\ 
\multirow{8}{*}{\textbf{40\%}} & MICE PMM & 1.82 & 0 & 35.0 & 1.94 & 0 & 38.1 & 5.22 & 0 & 43.0 \\ 
&  pan Fixed & 0.05 & 100 & 13.4 & 0.08 & \textbf{100} & 12.8 & 1.02 & 76 & 74.9 \\ 
&  pan Random & 0.13 & \textbf{99} & 21.0 & 0.20 & \textbf{100} & 31.4 & 4.73 & 0 & 93.7 \\ 
&  Amelia TS & 1.34 & 0 & 32.4 & 1.48 & 0 & 35.5 & 4.77 & 0 & 40.5 \\ 
&  Amelia CS & 0.69 & 3 & 24.6 & 0.86 & 0 & 24.6 & 6.04 & 0 & 69.1 \\ 
&  Amelia TSCS & 0.22 & \textbf{99} & 22.4 & 0.43 & 74 & 32.4 & 6.42 & 0 & 71.1 \\ 
 & MINTS & \textbf{0.04} & 100 & \textbf{4.6} & \textbf{0.07} & \textbf{100} & \textbf{5.4} & 0.63 & 13 & \textbf{31.9} \\ 
 & MINTS DA & 0.27 & 100 &  & 0.28 & \textbf{100} &  & \textbf{0.53} & \textbf{100} &  \\ 
 & &  & &  &  &  &  &  &  &  \\ 
\multirow{7}{*}{\textbf{80\%}} & MICE PMM & 2.96 & 0 & 52.1 & 3.08 & 0 & 54.2 &  &  &  \\ 
&  pan Fixed & 0.07 & \textbf{100} & 41.4 & \textbf{0.09} & \textbf{100} & 45.9 &  &  &  \\ 
&  pan Random & 1.20 & 0 & 69.3 & 1.29 & 0 & 76.6 &  &  &  \\ 
&  Amelia TS & 2.55 & 0 & 44.6 & 2.64 & 0 & 47.0 &  &  &  \\ 
&  Amelia CS & 1.45 & 0 & 57.8 & 1.52 & 0 & 57.3 &  &  &  \\ 
&  Amelia TSCS & 1.01 & 2 & 62.4 & 1.57 & 0 & 64.3 &  &  &  \\ 
&  MINTS & \textbf{0.06} & \textbf{100} & \textbf{21.8} & 0.11 & \textbf{100} & \textbf{25.5} &  &  &  \\ 
& MINTS DA & 0.53 & \textbf{100} &  & 0.54 & \textbf{100} &  &  &  &  \\ 
   \hline
\end{tabular}
}
\end{table}

We also considered a validation exercise using a random intercept analysis model for the regression of TFR on NER. 
For the 10\% rate of simulated missingness, MINTS has the best performance for estimation of the fixed effect coefficient on NER in the random intercept analysis model.
The pan Random Effects model tends to perform the best for estimation of the fixed effect coefficient at higher rates of missingness, while MINTS performs the best for estimation of the variance of the random intercepts.
Full details of this analysis model validation exercise can be found in Section 4 of the Supplementary Material (\cite{LiuRaftery2025}).

\subsection{Out-of-Sample Validation}
Next, we evaluated the predictive performance of each method for imputing the individual values of NER that were simulated as missing. 
In each experiment, the performance of point estimates was evaluated using the mean absolute error, where the mean was taken over all country-years in the testing set. 
The performance of the predictive intervals was evaluated by checking the average interval widths and coverage of the 95\% intervals with respect to the observations in the testing set, where coverage is calculated as the proportion of the intervals that contained the true left-out values of NER.
We also evaluated how well each imputation method balances the trade-off between interval width and coverage for the 95\% intervals using the average negatively-oriented interval score from \cite{GneitingRaftery2007}.
For a 95\% prediction interval, the interval score is defined as
\begin{align*}
IS =& \frac{1}{n} \sum_x \left[ (u - l) + \frac{2}{0.05}  1\{x < l\} + \frac{2}{0.05} 1\{x > u\} \right],
\end{align*}
where $(l, u)$ is the prediction interval, $n$ is the number of observations in the testing set, and the sum is over all true values $x$ for observations in the testing set. 

The out-of-sample validation exercise evaluates the posterior predictive distribution of the missing values given the observed values.
Samples from the posterior predictive distribution of the missing values from the MINTS method were obtained using a slight modification of the imputation procedure, details of which can be found in Section 7 of the Supplementary Material (\cite{LiuRaftery2025}).
Details of how samples from the posterior predictive distributions were obtained for the MICE and Amelia methods can also be found in Section 7 of the Supplementary Material (\cite{LiuRaftery2025}). 
Medians from the estimated posterior predictive distributions were used as point estimates for the imputed values and the 0.025 and 0.975 quantiles of the estimated posterior predictive distributions were used as 95\% interval estimates. 

\subsubsection{Results}
The results of the out-of-sample validation exercise are summarized in Table \ref{tab:val1}. 
MINTS results in the smallest MAE, narrowest interval width, and smallest interval score in all experiments.
MINTS has close to nominal coverage in all experiments except for MNAR 40\%, where coverage is below nominal.
MINTS generally also produces narrower interval widths compared to the existing methods; these intervals appear to be sufficiently wide under MCAR and MAR, but may be too narrow under MNAR. 

MICE PMM performs the worst overall, with comparatively large MAE and undercoverage in all experiments.
This is perhaps unsurprising, as MICE PMM is the only method that does not account for the hierarchical structure of the data. 
Amelia TS suffers from similarly large MAE as MICE PMM, but retains close to nominal coverage in experiments under MCAR and MAR thanks to its wider intervals. 
Out of the previously existing methods, pan Random Effects and Amelia TSCS perform the best overall.
Amelia TSCS tends towards undercoverage, having larger MAE and narrower intervals than pan Random Effects in most experiments, while pan Random Effects tends towards overcoverage. 

\begin{table}[!t]
\caption{Summary of out-of-sample validation for enrollment data for the country-years where NER was simulated as missing. MAE denotes mean absolute error, Cvg denotes the average coverage of 95\% intervals as a percentage, Width denotes the average width of 95\% intervals, and IS denotes the interval score for 95\% intervals. Results are averaged over all NER observations simulated as missing in each experiment.} \label{tab:val1}
\centerline{
\small
\begin{tabular}{cc|rrrr|rrrr|rrrr}
  \hline
Simulated  & \multirow{2}{*}{Method}  & \multicolumn{4}{c|}{\textbf{MCAR}}  & \multicolumn{4}{c|}{\textbf{MAR}} & \multicolumn{4}{c}{\textbf{MNAR}} \\
 \multicolumn{1}{c}{Missingness Rate} &  & \multicolumn{1}{c}{MAE} & \multicolumn{1}{c}{Cvg} & \multicolumn{1}{c}{Width} & \multicolumn{1}{c|}{IS} & \multicolumn{1}{c}{MAE} & \multicolumn{1}{c}{Cvg} & \multicolumn{1}{c}{Width} & \multicolumn{1}{c|}{IS} & \multicolumn{1}{c}{MAE} & \multicolumn{1}{c}{Cvg} & \multicolumn{1}{c}{Width} & \multicolumn{1}{c}{IS} \\
   \hline 
\multirow{8}{*}{\textbf{10\%}}  & MICE PMM & 5.79 & 89.6 & 22.7 & 38.1 & 7.06 & 86.7 & 22.9 & 60.4 & 12.66 & 74.1 & 30.9 & 104.7 \\ 
&  pan Fixed & 5.76 & 96.8 & 29.2 & 31.8 & 5.96 & 95.7 & 31.6 & 39.5 & 15.21 & 82.0 & 42.4 & 76.6 \\ 
&  pan Random & 2.02 & 98.6 & 19.4 & 20.5 & 2.33 & 98.6 & 20.3 & 21.0 & 3.28 & 98.9 & 30.1 & 37.0 \\ 
 & Amelia TS & 7.29 & 96.8 & 34.4 & 35.6 & 7.13 & \textbf{95.0} & 34.5 & 40.2 & 18.03 & 76.3 & 44.4 & 110.8 \\ 
&  Amelia CS & 3.50 & \textbf{95.3} & 19.8 & 27.7 & 4.29 & 91.7 & 19.2 & 31.7 & 8.71 & 87.4 & 29.4 & 125.9 \\ 
&  Amelia TSCS & 2.18 & 95.7 & 13.4 & 16.6 & 2.33 & 93.9 & 13.3 & 17.3 & 6.07 & 90.6 & 30.9 & 133.6 \\ 
&  MINTS & \textbf{1.27} & 96.8 & \textbf{9.7} & \textbf{11.7} & \textbf{1.42} & \textbf{95.0} & \textbf{9.6} & \textbf{12.3} & \textbf{1.77} & \textbf{94.6} & \textbf{8.5} & \textbf{13.1} \\ 
 & & & &  &  &  &  &  &  &  \\ 
\multirow{8}{*}{\textbf{40\%}} &  MICE PMM & 11.03 & 88.3 & 39.0 & 56.0 & 12.67 & 87.8 & 37.8 & 73.3 & 29.73 & 41.8 & 38.3 & 493.8 \\ 
&  pan Fixed & 10.08 & 97.7 & 47.1 & 51.0 & 9.43 & 97.8 & 46.7 & 50.3 & 31.06 & 28.5 & 36.5 & 616.2 \\ 
&  pan Random & 3.14 & 99.7 & 30.5 & 31.0 & 3.39 & 99.5 & 31.1 & 32.9 & 17.11 & \textbf{97.3} & 60.3 & 70.9 \\ 
&  Amelia TS & 12.45 & \textbf{93.9} & 49.1 & 57.6 & 11.25 & \textbf{94.6} & 46.6 & 52.9 & 32.54 & 28.7 & 37.2 & 637.9 \\ 
&  Amelia CS & 6.28 & 91.7 & 28.8 & 48.3 & 6.68 & 87.1 & 25.2 & 67.5 & 28.83 & 36.7 & 31.8 & 705.9 \\ 
&  Amelia TSCS & 3.28 & 92.1 & 15.8 & 31.5 & 3.89 & 90.5 & 16.3 & 52.2 & 27.28 & 41.3 & 24.1 & 818.4 \\ 
&  MINTS & \textbf{2.08} & \textbf{96.1} & \textbf{12.9} & \textbf{14.8} & \textbf{2.17} & \textbf{95.4} & \textbf{12.5} & \textbf{16.0} & \textbf{4.44} & 88.1 & \textbf{16.2} &\textbf{33.2} \\ 
 & &  & &  &  &  &  &  &  &  \\ 
\multirow{7}{*}{\textbf{80\%}} & MICE PMM & 18.36 & 89.8 & 61.5 & 87.9 & 20.14 & 88.7 & 62.0 & 92.8 & & & \\ 
&  pan Fixed & 16.89 & 97.6 & 71.6 & 76.0 & 16.80 & \textbf{95.4} & 69.5 & 77.6 & & & \\ 
&  pan Random & 5.49 & 99.9 & 60.5 & 60.6 & 6.08 & 100.0 & 58.6 & 58.6 & & & \\  
&  Amelia TS & 18.82 & 92.2 & 67.7 & 80.3 & 18.08 & 91.9 & 65.3 & 80.8 & & & \\  
&  Amelia CS & 10.32 & 88.7 & 40.5 & 90.9 & 10.49 & 77.0 & 31.5 & 153.3 & & & \\ 
&  Amelia TSCS & 6.40 & 82.9 & 25.0 & 114.1 & 8.47 & 74.4 & 24.2 & 165.1 & & & \\ 
&  MINTS & \textbf{4.14} & \textbf{93.7} & \textbf{20.1} & \textbf{24.8} & \textbf{4.86} & 95.5 & \textbf{23.1} & \textbf{28.3} & & & \\ 
   \hline
\end{tabular}
}
\end{table}

\begin{figure}[!hb]
	\centering
	\includegraphics[width = 0.65\textwidth]{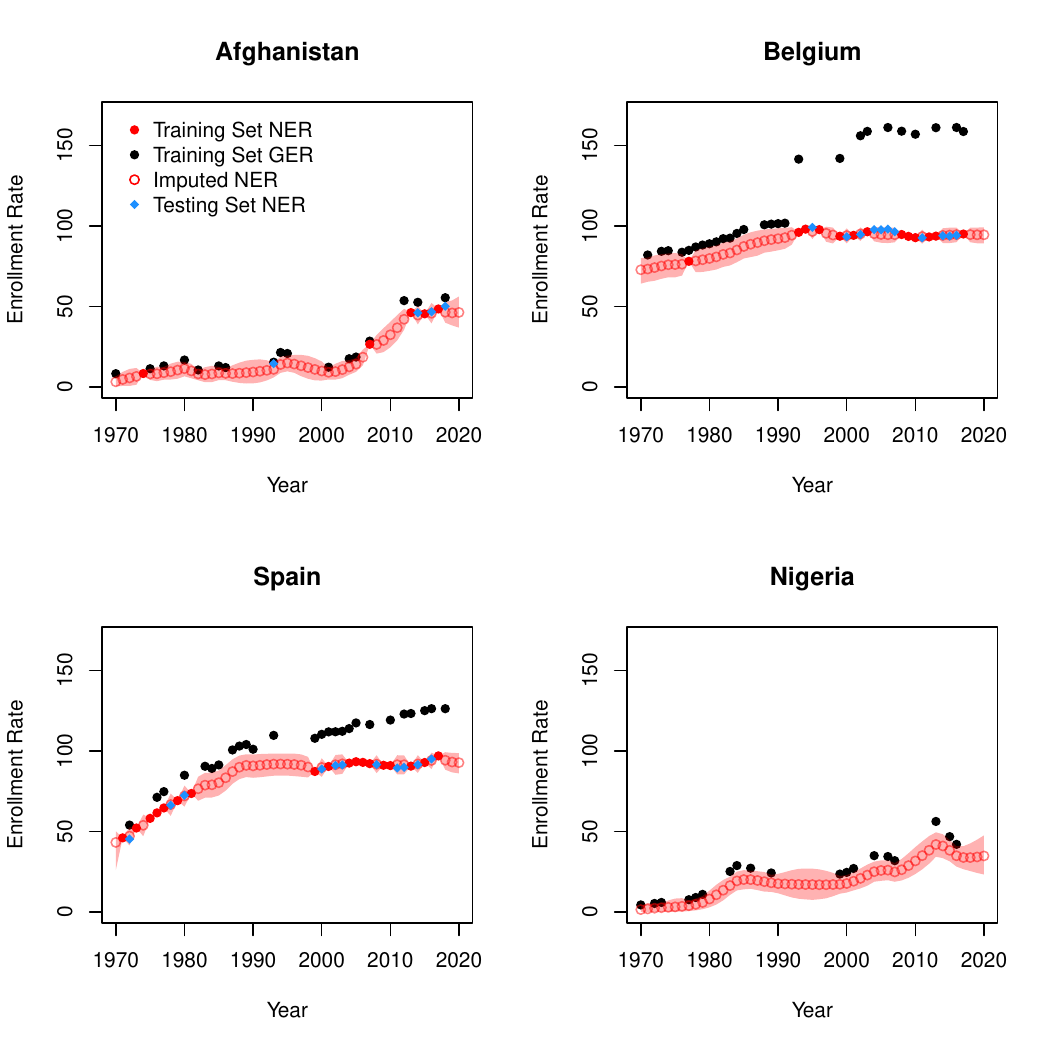}
	\caption{MCAR 40\% experiment results for selected countries from the out-of-sample validation for enrollment data. Solid black and red circles indicate values in the training set for GER and NER, respectively. Solid blue diamonds indicate the true values in the testing set for NER. Open red circles indicate the median imputed values of NER and the red shaded regions indicate the 95\% posterior quantiles for imputed values of NER, where values are imputed for country-years in the testing set and country-years that started as missing in the enrollment data set.} \label{fig:val1_MCAR40}
\end{figure}

\begin{figure}[!htb]
	\centering
	\includegraphics[width = 0.65\textwidth]{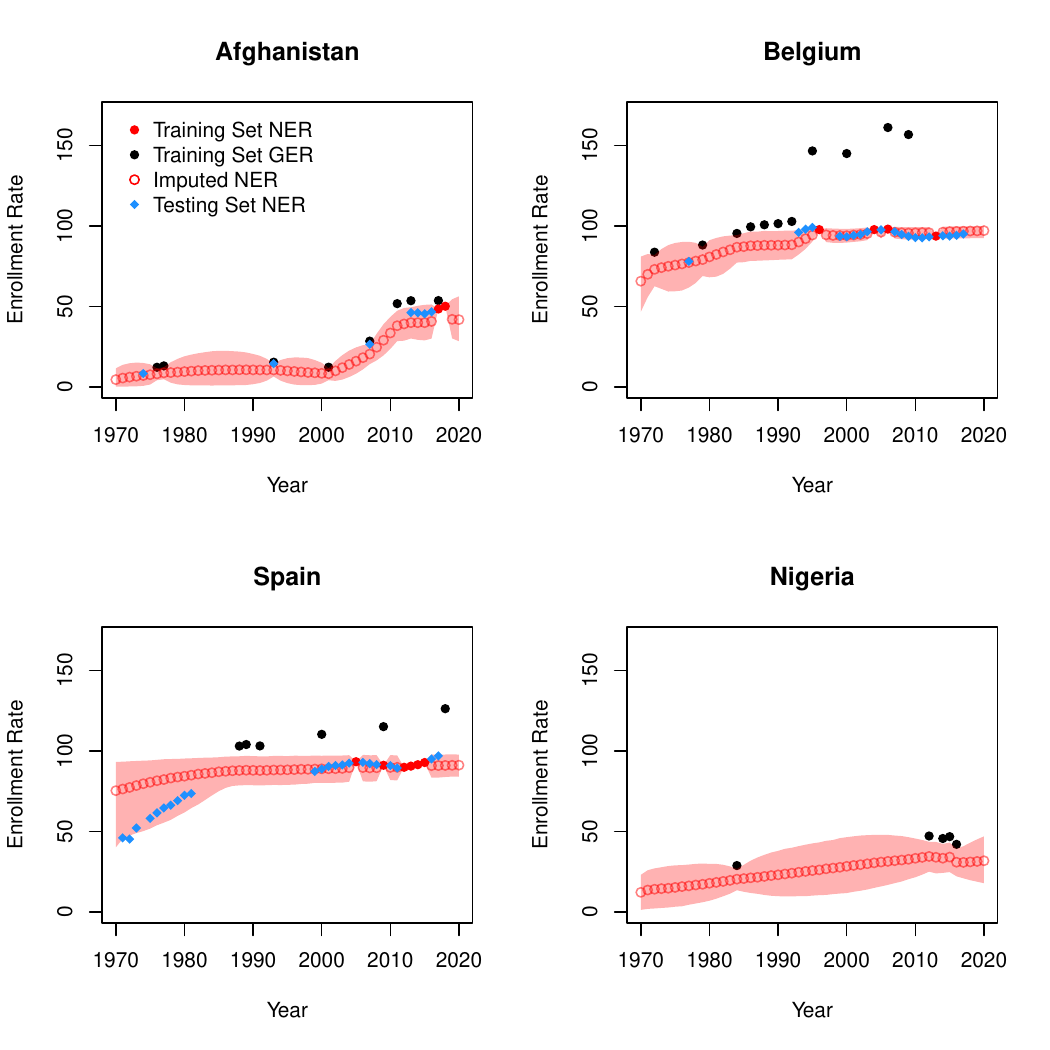}
	\caption{MAR 80\% experiment results for selected countries from the out-of-sample validation for enrollment data. Solid black and red circles indicate values in the training set for GER and NER, respectively. Solid blue diamonds indicate the true values in the testing set for NER. Open red circles indicate the median imputed values of NER and the red shaded regions indicate the 95\% posterior quantiles for imputed values of NER, where values are imputed for country-years in the testing set and country-years that started as missing in the enrollment data set.} \label{fig:val1_MAR80}
\end{figure}

We illustrate the out-of-sample validation results for the MINTS method for selected experiments in Figures \ref{fig:val1_MCAR40} and \ref{fig:val1_MAR80}.
Observations in the training set are shown as solid circles in black for GER and red for NER.
Posterior medians for the imputed values of NER are shown as red open circles, while the red shaded regions represent the 95\% posterior predictive intervals. 
Imputed values are plotted for both the country-years that started as missing in the school enrollment data set and the country-years that were simulated as missing for the testing set. 
The true values of NER for the country-years in the testing set are shown as solid blue diamonds. 
Figure \ref{fig:val1_MCAR40} shows the out-of-sample validation results for Afghanistan, Belgium, Spain, and Nigeria for the MCAR 40\% experiment.
Overall, the posterior predictive distributions for imputed values of NER result in plausible time series trends for each country, with the majority of the observations of NER that were simulated as missing captured within the 95\% intervals. 
Results for the same example countries are shown in Figure \ref{fig:val1_MAR80} for the MAR 80\% experiment. 
The predictive intervals are much wider overall for the MAR 80\% experiment compared to the MCAR 40\% experiment.
This is especially apparent for Spain, which has many observations of NER and GER in early years and thus had a high proportion of observed values that were simulated as missing.
Despite the very large amount of missing data in the experiments using the 80\% rate of simulated missingness, we find the MINTS still imputes plausible country-specific trends that reflect our knowledge of how school enrollment rates change over time.

\subsection{Full Enrollment Data Results} \label{sec:enroll_full}
We used MINTS to create 40 multiple imputations for the missing country-years in the original school enrollment data set without simulating any additional country-years as missing.
The original data set has a large amount of missing data for NER, with about 73.0\% of the 10,302 country-years missing.
However, we have partial information about the level of school enrollment arising from GER and the hierarchical time series structure of the data for many of these missing country-years.
About 48.8\% of the 7,524 country-years where NER is missing have observed values of GER.
In these country-years, the coarse measure of school enrollment (GER) provides us with partial information about the refined measure of school enrollment (NER).
Partial information about each country's likely level of enrollment and likely trend over time in enrollment can also be obtained from the hierarchical time series structure of the data.
For these reasons, we believe multiple imputation of NER is appropriate despite the large amount of missing data.

The prior distributions for MINTS were specified using the data-based algorithm to determine the value of the control parameters. 
For the global parameters, the data-based algorithm results in the following prior distributions:
\begin{align*}
    \sigma_X^2 \sim& InvGamma(2, \delta_X = 11.5), \\
    \mu_{drift} \sim& N(\nu_{drift} = 0.994 , \zeta_{drift}^2 = 0.00204), \\
    \sigma_{drift^2} \sim& InvGamma(2, \delta_{drift} = 1.87),  \\
    \sigma_{Y}^2 \sim& InvGamma(2, \delta_Y = 5.06), \\
    \mu_0 \sim& N(0, \zeta_0^2 = 0.00233), \\
    \sigma_0^2 \sim& InvGamma(2, \delta_0 = 2.65).
\end{align*}

The joint prior distribution of $(Y_{c,0}, X_{c,0})$ was estimated using an early subset of data that covered years 1970 to 1980 to balance between only including observations close to the first time period and having a sufficient number of observations in the chosen subset. 
The truncation for the joint prior distribution is specified based on substantive knowledge that $\mathbf{X}$ and $\mathbf{Y}$ are generally increasing over time.
The joint prior distribution of $(Y_{c,0}, X_{c,0})$ is
\begin{align*}
    \left[ \begin{matrix}
	Y_{c,0} \\
	X_{c,0}
	\end{matrix} \right] \sim& \, TN\left( \bm{\mu}_{early} = 
	\left[ \begin{matrix}
	40.4 \\
	47.2
	\end{matrix} \right] , 
	\bm{\Sigma}_{early} = 
	\left[ \begin{matrix}
	607.9  & \: 649.4 \\
	649.4  & \: 718.0 
	\end{matrix} \right] \right),
\end{align*}
where the truncation is such that $X_{c,0}$ is bounded from below by 0 and from above by the maximum observed value of GER for country $c$.
$Y_{c,0}$ is bounded from below by 0 and from above by the minimum of $X_{c,0}$ and the maximum observed value of NER for country $c$. 

To check that these prior distributions are appropriate for the school enrollment application, we conducted a prior predictive check and visually compared the prior and posterior distributions for the parameters specified using the data-based algorithm. 
We found that the chosen prior distributions are sufficiently diffuse and do not overwhelm the posterior. 
Details of these assessments can be found in Section 9 of the Supplementary Material (\cite{LiuRaftery2025}).

\begin{figure}[!hb]
	\centering
	\includegraphics[width = 0.65\textwidth]{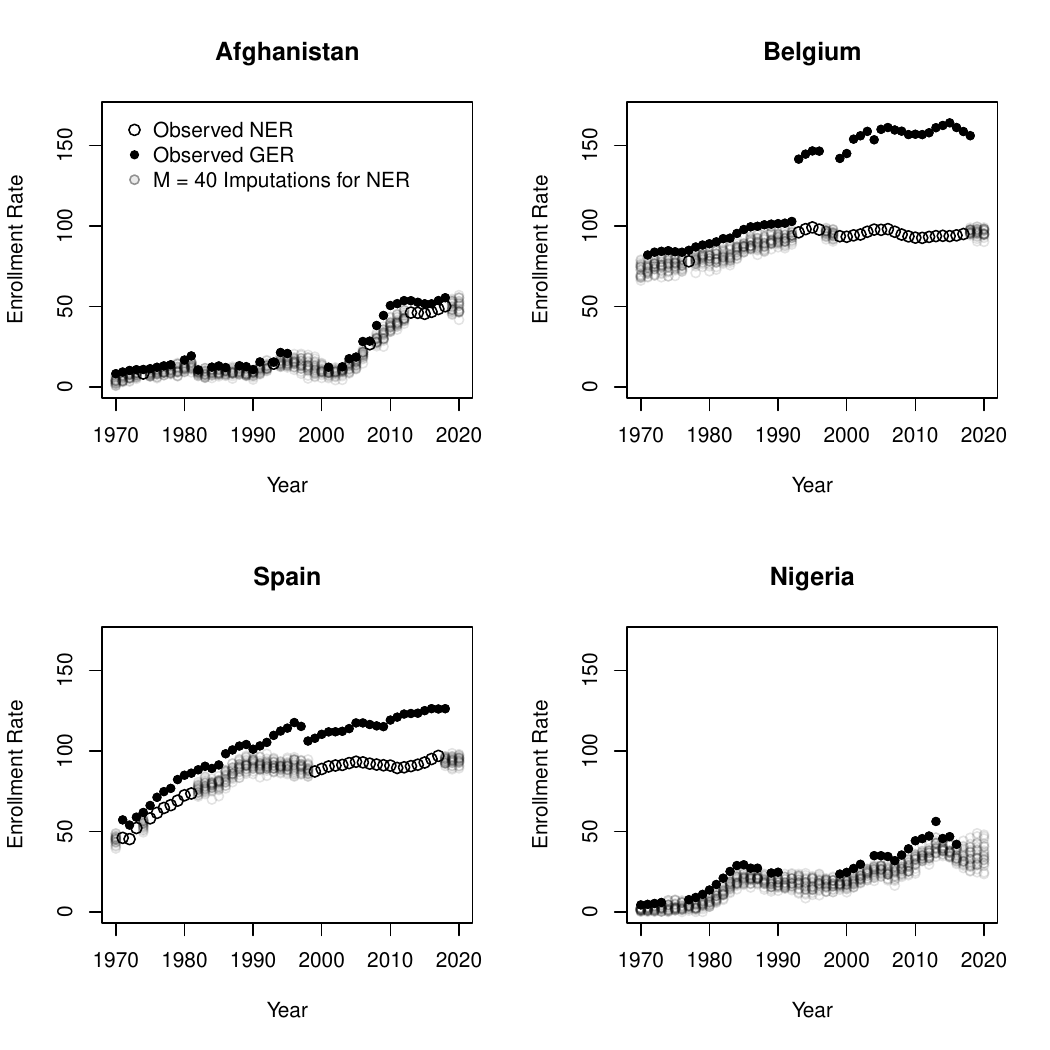}
	\caption{Results of $M = 40$ imputations for NER for selected countries. Open and solid black circles indicate observed values of NER and GER, respectively, from the enrollment data set. Translucent grey circles indicate imputed values of NER, where a total of 40 imputations were created for each missing value.} \label{fig:full}
\end{figure}

Figure \ref{fig:full} illustrates the multiply imputed values for NER from MINTS for Afghanistan, Belgium, Spain, and Nigeria. 
The multiple imputations for NER are shown as translucent grey circles, while the observed values of NER and GER from the original data set are shown as open and solid black circles, respectively. 
Despite the large amount of missing data for NER in the full enrollment data set, we find MINTS produces imputations that result in plausible imputed trends over time for all countries. 

A CSV file containing the 40 completed data sets for both NER and GER and a CSV file containing the median imputed values for both NER and GER are provided in the Supplementary Material (\cite{LiuRaftery2025}). 
We note that the intended use of MINTS is to create multiple imputations for the variable of interest NER, and the auxiliary variable GER is only imputed out of necessity. 
However, we make the imputed values of GER available to provide additional context for how the MINTS method works for the school enrollment application.

We also created multiply imputed estimates for the linear regression of TFR on NER, where $M = 40$ multiple imputations were created using MINTS and the existing multiple imputation methods considered in the validation exercises. 
We estimated the same linear regression of TFR on NER that was used in the analysis model validation in Section \ref{sec:enroll_analysis}, where the quantity of interest is the regression coefficient $\beta_1$.
Table \ref{tab:full_lm} summarizes the multiply imputed estimates of $\beta_1$ from MINTS and the six existing multiple imputation methods using the full enrollment data along with estimates from the data augmentation approach.

In the analysis model validation exercise using the school enrollment data, we found that MINTS resulted in the smallest MAE and closest to nominal coverage, and that pan Fixed Effects was the most competitive alternative out of the existing multiple imputation methods.
When applied to the full school enrollment data, MINTS and pan Fixed Effects result in similar multiply imputed estimates of $\beta_1$.
Both methods estimate that an increase in NER from 0\% to 100\% is associated with a reduction in TFR of over 4.2 children per woman, where MINTS estimates a smaller effect size than pan Fixed Effects.
MINTS and pan Fixed Effects also result in confidence intervals of approximately the same width, but pan Fixed Effects has a smaller estimated FMI.

\begin{table}[!htb]
\caption{Comparison of multiply imputed estimates of $Q = \beta_1$, the regression coefficient on NER in the linear regression of TFR on NER, created using $M = 40$ multiple imputations from MINTS and six existing multiple imputation methods. $\bar{Q}_{40}$ denotes the pooled point estimates, 95\% CI denotes the 95\% confidence intervals for $\bar{Q}_{40}$, and FMI denotes the fraction of missing information as a percentage. $\bar{Q}_{40}$ and the confidence interval bounds are multiplied by 100 before reporting.} \label{tab:full_lm}
\centerline{
\small
\begin{tabular}{c|rrr}
  \hline
Method & $\bar{Q}_{40}$ & 95\% CI & FMI \\
   \hline 
MICE PMM & -2.24 & (-2.40, -2.07) & 41.5 \\
pan Fixed & -4.36 & (-4.52, -4.21) &  39.7 \\
pan Random & -3.50 & (-3.86, -3.14) & 90.1 \\
Amelia TS & -2.85 & (-3.02, -2.68) & 34.0 \\
Amelia CS & -3.48  & (-3.67, -3.29) & 58.3\\
Amelia TSCS & -3.55  & (-3.79, -3.31) & 74.0 \\
MINTS & -4.26 & (-4.42, -4.11) & 53.2 \\
MINTS DA & -5.40 & (-5.73, -4.84) & \\
   \hline
\end{tabular}
}
\end{table}

The other five models based on existing methods (MICE PMM, pan Random Effects, Amelia TS, Amelia CS, and Amelia TSCS) had substantial bias and undercoverage in the analysis model validation exercise for estimation of $\beta_1$.
When applied to the full school enrollment data, these five models all lead to multiply imputed point estimates for $\beta_1$ that suggest an increase in NER from 0\% to 100\% is associated with a reduction in TFR of less than 3.6 children per woman. 
Based on the findings of the analysis model validation exercise, these estimates are likely to be biased compared to the estimates from MINTS and pan Fixed effects and may underestimate the relationship between TFR and NER. 

The data augmentation approach based on MINTS results in the largest estimated effect size, with a point estimate for $\beta_1$ that suggests an increase in NER from 0\% to 100\% is associated with a reduction in TFR of about 5.4 children per woman.
In the analysis model validation exercise for estimation of $\beta_1$, we found that the data augmentation approach had larger MAE than MINTS when the MAR assumption was satisfied.
However, the data augmentation approach had the smallest MAE and closest to nominal coverage in the MNAR experiments. 
If we have reason to suspect a strong violation of the MAR assumption for the full school enrollment data set, then the estimates from MINTS and pan Fixed Effects may also underestimate the true relationship between TFR and NER.

An important limitation of the multiple imputation results presented in this section is that the complete cases in the school enrollment data set only comprise 26.8\% of the possible country-years.
The MINTS method therefore relies on a relatively small proportion of the data to estimate the relationship between NER and GER.
Our validation exercises using the school enrollment data suggest that MINTS produces unbiased multiply imputed estimates with reasonable coverage even in settings with a very high percentage of missing data.
However, the small number of complete cases increases the risk that the assumptions of MINTS are violated.
The restriction of MINTS to the bivariate setting increases the risk that the ignorability assumption is violated, as the probability of NER being missing for the country-years where both NER and GER are missing may depend on external factors that are not accounted for in the bivariate imputation model.
The country-years with both NER and GER missing could also be systematically different from the country-years where at least one measure of school enrollment is observed, and the relationships between NER and GER estimated using the observed country-years may not be appropriate for country-years where both measures of enrollment are missing. 
If the true missing data mechanism is strongly MNAR, the results from our simulation studies and validation exercises suggest that multiply imputed point estimates from MINTS are likely to be biased and interval estimates are likely to suffer from undercoverage.
Incorporation of additional variables that are predictive of the probability of observing NER and GER into the imputation model could improve the plausibility of the MAR assumption, and generalization of the MINTS method to the multivariate setting to accommodate such variables is of interest for future work.

\section{Discussion}
We developed a multiple imputation method for hierarchical time series data that accommodates nonlinear relationships between variables.
We evaluated the performance of the proposed method using a simulation study and an application to a data set on secondary school enrollment rates. 
Through comparisons with existing methods for multiple imputation of hierarchical time series data, we found that the proposed MINTS method can lead to substantial gains in performance when variables in the imputation model have a nonlinear relationship. 

In the simulation study, MINTS generally performed better for estimation of parameters in uncongenial analysis models compared to existing models based on the MICE and Amelia methodologies. 
We found that MINTS is likely to balance the tradeoff between MAE and coverage better than existing methods when there is a nonlinear relationship between variables, though undercoverage is a concern at very high rates of missingness when the number of complete cases is small. 
MINTS still performed well in validation exercises with a linear relationship between variables, but we found the difference in performance between MINTS and the existing methods was less pronounced.

For the application to the school enrollment data, we evaluated how well MINTS performed for estimation of parameters in uncongenial analysis models and for prediction of individual missing values. 
MINTS resulted in improved performance for estimation of the linear regression of TFR on NER compared to existing models based on MICE and Amelia, with the smallest or second smallest MAE in all experiments and close to nominal coverage in all but one experiment. 
MINTS generally had good performance for estimation of the random intercept model regressing TFR on NER, with the smallest MAE for estimation of the variance of the random intercepts across experiments. 
However, pan Random Effects performed better for estimation of the fixed effect coefficient on NER at higher rates of missingness.
For random intercept models, MINTS is likely to be a better choice than existing methods if a components of variance analysis is of interest, but pan Random Effects could be a better choice if only the fixed effect coefficient is of interest.
In the out-of-sample validation exercise, we found that MINTS resulted in substantial improvements for prediction of individual missing values of NER compared to existing methods, with smaller MAE, narrower intervals, and smaller interval scores across experiments while maintaining close to nominal coverage in all but one experiment.

To facilitate easier comparisons across validation exercises, we summarize the average MAE across experiments for each multiple imputation model and the data augmentation approach based on MINTS in Table \ref{tab:avg_MAE}.  
A version of Table \ref{tab:avg_MAE} that includes all validation exercises conducted is available in Section 8 of the Supplementary Material (\cite{LiuRaftery2025}).
We note that Table \ref{tab:avg_MAE} should not be interpreted as inferential and is only of interest as an exploratory comparison tool to enable a simple comparison of one metric from the full evaluation results. 
Out of the multiple imputation methods considered, MINTS consistently results in the smallest average MAE across experiments in each validation exercise. 
For the validation exercise using the nonlinear simulated data, the data augmentation approach resulted in the smallest average MAE, followed by multiple imputation using MINTS. 

\begin{table}[!tb]
\caption{Average mean absolute error (MAE) across all experiments for each multiple imputation method within each validation exercise. For the enrollment data, OOS denotes to the out-of-sample validation and $\beta_1$ is the parameter from the linear regression analysis model validation. For the nonlinear simulated data, $\omega_1$ is the parameter from the linear regression analysis model validation. MAE for the $\beta_1$ and $\omega_1$ columns are multiplied by 100 before reporting.} \label{tab:avg_MAE}
\centerline{
\small
\begin{tabularx}{0.8\textwidth}{c| *{2}{Y}| *{1}{Y}}
  \hline
\multirow{2}{*}{Method}  & \multicolumn{2}{c|}{Enrollment Data} & \multicolumn{1}{c}{Nonlinear Data}  \\
 & \multicolumn{1}{c}{OOS} & \multicolumn{1}{c|}{$\beta_1$} & \multicolumn{1}{c}{$\omega_1$} \\
   \hline 
MICE PMM & 14.68 & 2.97  & 18.28 \\
pan Fixed & 13.90 & 0.52 & 14.37   \\
pan Random & 5.36 & 1.48  & 16.15  \\
Amelia TS & 15.70 & 2.59 & 12.07  \\
Amelia CS & 9.89 & 2.15  & 29.86  \\
Amelia TSCS & 7.49 & 2.04 & 12.93\\
MINTS & \textbf{2.77} & \textbf{0.18} & 4.15 \\
MINTS DA &  & 0.38 & \textbf{2.33} \\
\hline
\end{tabularx}
}
\end{table}

We note that all of the multiple imputation methods compared assume that the missing data mechanism is ignorable. 
We conducted sensitivity analyses to evaluate how robust each imputation model is to one type of violation of the ignorability assumption through the MNAR experiments. 
In the simulation studies, we found that all imputation methods considered were relatively robust to the violation of ignorability when the rate of simulated missingness was low.
However, substantial increases in bias and undercoverage were seen at the 40\% and 80\% rates. 
MINTS tended to be the most robust to these violations but still suffered from bias and undercoverage, especially at the 80\% rate. 
No imputation method performed consistently well in the MNAR experiments for the school enrollment data, though we found MINTS and the pan models tended to be the most robust to violations of ignorability.
Across validation exercises, the data augmentation approach based on MINTS generally performed better than all of the multiple imputation methods in the MNAR experiments.

Our results suggest that MINTS is generally able to balance good predictive performance for individual missing values with good performance for multiply-imputed estimates of parameters in substantive analysis models.
However, MINTS should be used with caution if a strong violation of the MAR assumption is suspected.
Additionally, MINTS is not recommended for use if there is a very small number of observed values available to estimate the imputation model.
This small sample size setting occurred for the nonlinear and linear simulated data sets, where some replications of the experiments at the 80\% rate of simulated missingness had fewer than 15 complete cases of $\mathbf{X}$ and $\mathbf{Y}$ after simulating data as missing. 
Although our validation exercises suggest MINTS tends to perform better than the existing multiple imputation methods in this small sample size setting, MINTS still resulted in larger bias than the data augmentation approach and had below nominal coverage.
If analysts have access to $\mathbf{X}$, $\mathbf{Y}$, and $\mathbf{Z}$, data augmentation could be a better alternative to multiple imputation in settings with a strong violation of the MAR assumption or with a very small number of complete cases. 
However, the data augmentation approach may not be practical if analysts are interested in more than one analysis using $\mathbf{Y}$. 
Unlike multiple imputation, the data augmentation approach is specific to a given analysis model and must be run separately for each analysis model of interest.

In this paper, we focused on the setting where the analysis model is uncongenial to the multiple imputation model.
This is a common occurrence for social science data, particularly when the imputation phase is conducted independently from the analysis phase. 
However, if the outcome variable $\mathbf{Z}$ is known during the imputation phase, congeniality between analysis and imputation models is a worthwhile goal. 
Several methods have been proposed for multiple imputation of hierarchical data under congeniality where the imputation model incorporates features of the substantive analysis model, such as \cite{Goldstein2014, EndersDuKeller2020, Ludtke2020, GrundLudtkeRobitzsch2021}; among others.
Many of these methods can account for nonlinear relationships, either through user-specified functional forms (e.g. \cite{Erler2019}) or through automated modeling procedures (e.g. \cite{Kim2014}).
Of particular note, \cite{Ludtke2020} and \cite{GrundLudtkeRobitzsch2021} propose a method called ``mdmb'' that uses a sequential decomposition of the joint model. 
The mdmb method ensures congeniality between analysis and imputation models by incorporating a term representing the substantive analysis model into the sequential decomposition following the substantive-model-compatible philosophy of \cite{Bartlett2015}. 
MINTS could be extended to congeniality using the same strategy as mdmb by adding a term representing the substantive analysis model into the decomposition of the joint model as
\begin{align*}
p(\mathbf{Z}, \mathbf{Y}, \mathbf{X} | \bm{\theta}_Z, \bm{\theta}) =& p(\mathbf{Z}| \mathbf{Y}, \mathbf{X}, \bm{\theta}_Z) p(\mathbf{Y} | \mathbf{X}, \bm{\theta}_Y) p(\mathbf{X} | \bm{\theta}_X).
\end{align*}
When $p(\mathbf{Z}| \mathbf{Y}, \mathbf{X}, \bm{\theta}_Z)$ is the exact substantive analysis model of interest, this decomposition corresponds to the data augmentation approach considered in our validation exercises.

\section{Conclusion}
We have proposed the MINTS method for multiple imputation of hierarchical nonlinear time series data.
We considered the bivariate setting where one variable is the variable of interest for analysis that represents a refined measure of an underlying quantity of interest, and the second variable is an auxiliary variable that represents a coarse measure of the same underlying quantity and has a nonlinear relationship with the variable of interest. 
The refined measure is more difficult to measure and has a greater amount of missing data than the coarse measure. 
This setting is motivated by a secondary school enrollment data set that includes two measures of school enrollment rates.
The Net Enrollment Rate (NER) is the variable of interest for an analysis of the relationship between the Total Fertility Rate and NER. 
Measurement of NER can be difficult, as NER incorporates information about both the number and ages of children enrolled in school.
The Gross Enrollment Ratio (GER) is a second measure school enrollment that only requires knowledge of the number of children enrolled in school.
GER is easier to measure and has a smaller amount of missing data than NER.
The MINTS method leverages the strong nonlinear relationship between NER and GER to impute missing values in NER using a combination of smoothing splines, country-specific intercepts, and time series methods.  

We compared MINTS with several existing methods for multiple imputation of hierarchical time series data through a simulation study and an application to the school enrollment data. 
We considered three models within the MICE framework of \cite{vanBuurenOudshoorn2011} and three models within the Amelia II framework of \cite{HonakerKing2010}. 
We found that MINTS generally resulted in better performance for estimation of parameters in uncongenial linear regression and random intercept models compared to imputation models based on the MICE and Amelia methodologies in both a simulation study using nonlinear simulated data and an application to the school enrollment data, though undercoverage can be a concern when the sample size is small. 
We also conducted an out-of-sample validation exercise for prediction of individual missing values using the school enrollment data and found that MINTS resulted in better predictive performance compared to the existing methods.
Based on these validation exercises, we believe MINTS has a promising capability to improve the quality of multiply imputed data sets for hierarchical nonlinear time series data.

One limitation of the MINTS method is the use of spline estimation to model the nonlinear relationship between variables in the imputation model.
The A-splines method of \cite{Goepp2018} helps to automate the estimation procedure, but as with all spline estimation methodologies there is the risk of overfitting. 
The MINTS algorithm uses linear splines and modifies the number of starting knots used in the A-spline algorithm when the sample size is small to reduce the risk of overfitting, but the spline estimation settings are likely to require manual tuning for applications of MINTS to other data sets. 
The risk of overfitting is especially of concern when the number of complete cases is small. 
The splines $f$ and $h$ estimated using A-splines should be checked for each application to ensure the estimated relationships make sense based on substantive knowledge of the variables in the imputation model. 
We note that other curve-fitting methods, such as LOESS, could be used instead. 
The estimation of $f$ and $h$ in MINTS is also limited by the assumption that the nonlinear relationship between $\mathbf{X}$ and $\mathbf{Y}$ does not change substantially across countries and times.
If there are a sufficient number of complete cases in the data set, country- or time-specific estimates of $f$ and $h$ could be used instead.  

The illustrations of MINTS presented in the simulation studies and the application to the school enrollment data are additionally limited by the use of a data-based algorithm to specify certain hyperparameters in the prior distributions. 
Although we found that the resulting prior distributions were sufficiently diffuse and did not overwhelm the posterior for these specific applications, a consequence of using the data-based algorithm is that the posterior distributions used to create multiple imputations are approximate rather than fully Bayesian posterior distributions. 
Alternative methods for specifying these hyperparameters that do not rely on the data, such as expert elicitation or using information from previous studies, could be used to address this limitation.

MINTS also has several practical limitations compared to MICE and Amelia.
Currently, MINTS is restricted to the bivariate setting with continuous variables.
While we are primarily motivated by multiple imputation in settings like the school enrollment rates application with two highly related variables that aim to measure the same underlying quantity, incorporation of additional variables into the MINTS model could improve the plausibility of the MAR assumption. 
In principle, MINTS could be extended to the multivariate setting by adding additional univariate conditional terms to the sequential decomposition of the joint distribution, where careful consideration is needed for the ordering of the added conditional distributions.
The imputation models for the added variables could be assumed to be linear or could incorporate a separate spline term for each marginal relationship in the conditional imputation model.
MINTS could also be extended to accommodate categorical variables through the use of generalized regression models, for example using similar methodology as \cite{LeeMitra2016}.
Another practical limitation of MINTS is computation time, where the estimation of the MINTS model is much more computationally intense than estimation of the MICE models and the two simpler Amelia models. 
Computation time could be improved by coding the MCMC algorithm in a more efficient language than R. 
Extending the MINTS methodology to the multivariate and mixed data type setting and improving the computational efficiency of the MINTS sampling algorithm is of interest for future work.

\begin{acks}[Acknowledgments]
The authors would like to thank the members of the working group on Applied, Bayesian, and Computational Statistics at the University of Washington for their helpful discussion. We also thank the referees, Associate Editor, and Editor for their valuable feedback. 
\end{acks}

\begin{funding}
This work was supported by NICHD grant R01 HD070936. 
\end{funding}

\begin{supplement}
\stitle{Supplementary Material}
\sdescription{PDF containing additional methodological details and validation results}
\end{supplement}

\begin{supplement}
\stitle{MINTS\_enrollment\_results.zip}
\sdescription{ZIP file containing a CSV of 40 multiple imputations for secondary school enrollment rates created using the MINTS method, a CSV of the medians of the multiply imputed values, and R code to reproduce the multiple imputations}
\end{supplement}


\bibliographystyle{imsart-nameyear} 
\bibliography{paper3}  

\begin{thebibliography}{62}

\bibitem[\protect\citeauthoryear{Axinn and Barber}{2001}]{AxinnBarber2001}
\begin{barticle}[author]
\bauthor{\bsnm{Axinn},~\bfnm{William~G.}\binits{W.~G.}} \AND
  \bauthor{\bsnm{Barber},~\bfnm{Jennifer~S.}\binits{J.~S.}}
(\byear{2001}).
\btitle{Mass Education and Fertility Transition}.
\bjournal{American Sociological Review}
\bvolume{66}
\bpages{481--505}.
\bdoi{10.2307/3088919}
\end{barticle}
\endbibitem


\bibitem[\protect\citeauthoryear{Bartlett et~al.}{2015}]{Bartlett2015}
\begin{barticle}[author]
\bauthor{\bsnm{Bartlett},~\bfnm{Jonathan~W.}\binits{J.~W.}},
  \bauthor{\bsnm{Seaman},~\bfnm{Shaun~R.}\binits{S.~R.}},
  \bauthor{\bsnm{White},~\bfnm{Ian~R.}\binits{I.~R.}} \AND
  \bauthor{\bsnm{Carpenter},~\bfnm{James~R.}\binits{J.~R.}}
(\byear{2015}).
\btitle{Multiple imputation of covariates by fully conditional specification:
  accommodating the substantive model}.
\bjournal{Statistical Methods in Medical Research}
\bvolume{24}
\bpages{462--487}.
\bdoi{10.1177/0962280214521348}
\end{barticle}
\endbibitem

\bibitem[\protect\citeauthoryear{Bodner}{2008}]{Bodner2008}
\begin{barticle}[author]
\bauthor{\bsnm{Bodner},~\bfnm{Todd~E.}\binits{T.~E.}}
(\byear{2008}).
\btitle{What Improves with Increased Missing Data Imputations?}
\bjournal{Structural Equation Modeling: A Multidisciplinary Journal}
\bvolume{15}
\bpages{651--675}.
\bdoi{10.1080/10705510802339072}
\end{barticle}
\endbibitem

\bibitem[\protect\citeauthoryear{Burgette and
  Reiter}{2010}]{BurgetteReiter2010}
\begin{barticle}[author]
\bauthor{\bsnm{Burgette},~\bfnm{Lane~F.}\binits{L.~F.}} \AND
  \bauthor{\bsnm{Reiter},~\bfnm{Jerome~P.}\binits{J.~P.}}
(\byear{2010}).
\btitle{Multiple Imputation for Missing Data via Sequential Regression Trees}.
\bjournal{American Journal of Epidemiology}
\bvolume{172}
\bpages{1070–-1076}.
\bdoi{10.1093/aje/kwq260}
\end{barticle}
\endbibitem

\bibitem[\protect\citeauthoryear{Caldwell}{1982}]{Caldwell1982}
\begin{bbook}[author]
\bauthor{\bsnm{Caldwell},~\bfnm{John~C.}\binits{J.~C.}}
(\byear{1982}).
\btitle{Theory of Fertility Decline}.
\bpublisher{Academic Press}, \baddress{New York}.
\end{bbook}
\endbibitem

\bibitem[\protect\citeauthoryear{Caldwell, Reddy and
  Caldwell}{1985}]{Caldwell1985}
\begin{barticle}[author]
\bauthor{\bsnm{Caldwell},~\bfnm{John~C.}\binits{J.~C.}},
  \bauthor{\bsnm{Reddy},~\bfnm{P.~H.}\binits{P.~H.}} \AND
  \bauthor{\bsnm{Caldwell},~\bfnm{Pat}\binits{P.}}
(\byear{1985}).
\btitle{Educational Transition in Rural South {I}ndia}.
\bjournal{Population and Development Review}
\bvolume{11}
\bpages{29--51}.
\bdoi{10.2307/1973377}
\end{barticle}
\endbibitem

\bibitem[\protect\citeauthoryear{Darnieder}{2011}]{Darnieder2011}
\begin{barticle}[author]
\bauthor{\bsnm{Darnieder},~\bfnm{William~F.}\binits{W.~F.}}
(\byear{2011}).
\btitle{Bayesian Methods for Data-Dependent Priors}.
\bjournal{Doctoral dissertation, Ohio State University}.
\bnote{OhioLINK Electronic Theses and Dissertations Center.
  \url{http://rave.ohiolink.edu/etdc/view?acc_num=osu1306344172}}.
\end{barticle}
\endbibitem

\bibitem[\protect\citeauthoryear{de~Jong, van Buuren and
  Spiess}{2016}]{deJong2016}
\begin{barticle}[author]
\bauthor{\bparticle{de} \bsnm{Jong},~\bfnm{Roel}\binits{R.}},
  \bauthor{\bparticle{van} \bsnm{Buuren},~\bfnm{Stef}\binits{S.}} \AND
  \bauthor{\bsnm{Spiess},~\bfnm{Martin}\binits{M.}}
(\byear{2016}).
\btitle{Multiple Imputation of Predictor Variables Using Generalized Additive
  Models}.
\bjournal{Communications in Statistics - Simulation and Computation}
\bvolume{45}
\bpages{1–-18}.
\bdoi{10.1080/03610918.2014.911894}
\end{barticle}
\endbibitem

\bibitem[\protect\citeauthoryear{Easterlin and
  Crimmins}{1985}]{EasterlinCrimmins1985}
\begin{bbook}[author]
\bauthor{\bsnm{Easterlin},~\bfnm{Richard~A.}\binits{R.~A.}} \AND
  \bauthor{\bsnm{Crimmins},~\bfnm{Eileen~M.}\binits{E.~M.}}
(\byear{1985}).
\btitle{The Fertility Revolution: A Supply-Demand Analysis}.
\bpublisher{University of Chicago Press}, \baddress{Chicago}.
\end{bbook}
\endbibitem

\bibitem[\protect\citeauthoryear{Edwards, Lindman and
  Savage}{1963}]{EdwardsLindmanSavage1963}
\begin{barticle}[author]
\bauthor{\bsnm{Edwards},~\bfnm{Ward}\binits{W.}},
  \bauthor{\bsnm{Lindman},~\bfnm{Harold}\binits{H.}} \AND
  \bauthor{\bsnm{Savage},~\bfnm{Leonard~J}\binits{L.~J.}}
(\byear{1963}).
\btitle{Bayesian statistical inference for psychological research.}
\bjournal{Psychological review}
\bvolume{70}
\bpages{193–242}.
\bdoi{10.1037/h0044139}
\end{barticle}
\endbibitem

\bibitem[\protect\citeauthoryear{Enders}{2010}]{Enders2010}
\begin{bbook}[author]
\bauthor{\bsnm{Enders},~\bfnm{Craig~K.}\binits{C.~K.}}
(\byear{2010}).
\btitle{Applied Missing Data Analysis}.
\bpublisher{Guilford Press}.
\end{bbook}
\endbibitem

\bibitem[\protect\citeauthoryear{Enders, Du and
  Keller}{2020}]{EndersDuKeller2020}
\begin{barticle}[author]
\bauthor{\bsnm{Enders},~\bfnm{Craig~K.}\binits{C.~K.}},
  \bauthor{\bsnm{Du},~\bfnm{Han}\binits{H.}} \AND
  \bauthor{\bsnm{Keller},~\bfnm{Brian~T.}\binits{B.~T.}}
(\byear{2020}).
\btitle{A Model-Based Imputation Procedure for Multilevel Regression Models
  With Random Coefficients, Interaction Effects, and Nonlinear Terms}.
\bjournal{Psychological Methods}
\bvolume{25}
\bpages{88--112}.
\bdoi{10.1037/met0000228}
\end{barticle}
\endbibitem

\bibitem[\protect\citeauthoryear{Enders, Mistler and
  Keller}{2016}]{EndersMistlerKeller2016}
\begin{barticle}[author]
\bauthor{\bsnm{Enders},~\bfnm{Craig~K.}\binits{C.~K.}},
  \bauthor{\bsnm{Mistler},~\bfnm{Stephen~A.}\binits{S.~A.}} \AND
  \bauthor{\bsnm{Keller},~\bfnm{Brian~T.}\binits{B.~T.}}
(\byear{2016}).
\btitle{Multilevel Multiple Imputation: A Review and Evaluation of Joint
  Modeling and Chained Equations Imputation}.
\bjournal{Psychological Methods}
\bvolume{21}
\bpages{222--240}.
\bdoi{10.1037/met0000063}
\end{barticle}
\endbibitem

\bibitem[\protect\citeauthoryear{Erler et~al.}{2019}]{Erler2019}
\begin{barticle}[author]
\bauthor{\bsnm{Erler},~\bfnm{Nicole~S.}\binits{N.~S.}},
  \bauthor{\bsnm{Rizopoulos},~\bfnm{Dimitris}\binits{D.}},
  \bauthor{\bsnm{Jaddoe},~\bfnm{Vincent W.~V.}\binits{V.~W.~V.}},
  \bauthor{\bsnm{Franco},~\bfnm{Oscar~H.}\binits{O.~H.}} \AND
  \bauthor{\bsnm{Lesaffre},~\bfnm{Emmanuel M. E.~H.}\binits{E.~M. E.~H.}}
(\byear{2019}).
\btitle{Bayesian imputation of time-varying covariates in linear mixed models}.
\bjournal{Statistical Methods in Medical Research}
\bvolume{28}
\bpages{555–-568}.
\bdoi{10.1177/0962280217730851}
\end{barticle}
\endbibitem

\bibitem[\protect\citeauthoryear{Gelman, King and Liu}{1998}]{Gelman1998}
\begin{barticle}[author]
\bauthor{\bsnm{Gelman},~\bfnm{Andrew}\binits{A.}},
  \bauthor{\bsnm{King},~\bfnm{Gary}\binits{G.}} \AND
  \bauthor{\bsnm{Liu},~\bfnm{Chuanhai}\binits{C.}}
(\byear{1998}).
\btitle{Not Asked and Not Answered: Multiple Imputation for Multiple Surveys}.
\bjournal{Journal of the American Statistical Association}
\bvolume{93}
\bpages{846--857}.
\bdoi{10.1080/01621459.1998.10473737}
\end{barticle}
\endbibitem

\bibitem[\protect\citeauthoryear{Gelman and Rubin}{1992}]{GelmanRubin1992}
\begin{barticle}[author]
\bauthor{\bsnm{Gelman},~\bfnm{Andrew}\binits{A.}} \AND
  \bauthor{\bsnm{Rubin},~\bfnm{Donald~B.}\binits{D.~B.}}
(\byear{1992}).
\btitle{Inference from Iterative Simulation Using Multiple Sequences}.
\bjournal{Statist. Sci.}
\bvolume{7}
\bpages{457--472}.
\bdoi{10.1214/ss/1177011136}
\end{barticle}
\endbibitem

\bibitem[\protect\citeauthoryear{Gneiting and
  Raftery}{2007}]{GneitingRaftery2007}
\begin{barticle}[author]
\bauthor{\bsnm{Gneiting},~\bfnm{Tilmann}\binits{T.}} \AND
  \bauthor{\bsnm{Raftery},~\bfnm{Adrian~E.}\binits{A.~E.}}
(\byear{2007}).
\btitle{Strictly Proper Scoring Rules, Prediction, and Estimation}.
\bjournal{Journal of the American Statistical Association}
\bvolume{102}
\bpages{359--378}.
\bdoi{10.1198/016214506000001437}
\end{barticle}
\endbibitem

\bibitem[\protect\citeauthoryear{Goepp}{2022}]{aspline}
\begin{bmisc}[author]
\bauthor{\bsnm{Goepp},~\bfnm{Vivien}\binits{V.}}
(\byear{2022}).
\btitle{aspline: Spline Regression with Adaptive Knot Selection}.
\bnote{R package version 0.2.0}.
\end{bmisc}
\endbibitem

\bibitem[\protect\citeauthoryear{Goepp, Bouaziz and Nuel}{2018}]{Goepp2018}
\begin{barticle}[author]
\bauthor{\bsnm{Goepp},~\bfnm{Vivien}\binits{V.}},
  \bauthor{\bsnm{Bouaziz},~\bfnm{Olivier}\binits{O.}} \AND
  \bauthor{\bsnm{Nuel},~\bfnm{Gr{\'e}gory}\binits{G.}}
(\byear{2018}).
\btitle{Spline Regression with Automatic Knot Selection}.
\bjournal{arXiv preprint arXiv:1808.01770}.
\bdoi{10.48550/arXiv.1808.01770}
\end{barticle}
\endbibitem

\bibitem[\protect\citeauthoryear{Goldstein, Carpenter and
  Browne}{2014}]{Goldstein2014}
\begin{barticle}[author]
\bauthor{\bsnm{Goldstein},~\bfnm{Harvey}\binits{H.}},
  \bauthor{\bsnm{Carpenter},~\bfnm{James~R.}\binits{J.~R.}} \AND
  \bauthor{\bsnm{Browne},~\bfnm{William~J.}\binits{W.~J.}}
(\byear{2014}).
\btitle{Fitting multilevel multivariate models with missing data in responses
  and covariates that may include interactions and non-linear terms}.
\bjournal{Journal of the Royal Statistical Society Series A: Statistics in
  Society}
\bvolume{177}
\bpages{553--564}.
\bdoi{10.1111/rssa.12022}
\end{barticle}
\endbibitem

\bibitem[\protect\citeauthoryear{Grund, L{\"u}dtke and
  Robitzsch}{2021}]{GrundLudtkeRobitzsch2021}
\begin{barticle}[author]
\bauthor{\bsnm{Grund},~\bfnm{Simon}\binits{S.}},
  \bauthor{\bsnm{L{\"u}dtke},~\bfnm{Oliver}\binits{O.}} \AND
  \bauthor{\bsnm{Robitzsch},~\bfnm{Alexander}\binits{A.}}
(\byear{2021}).
\btitle{Multiple imputation of missing data in multilevel models with the R
  package mdmb: a flexible sequential modeling approach}.
\bjournal{Behavior Research Methods}
\bvolume{53}
\bpages{2631--2649}.
\bdoi{10.3758/s13428-020-01530-0}
\end{barticle}
\endbibitem

\bibitem[\protect\citeauthoryear{He, Yucel and
  Raghunathan}{2011}]{HeYucelRaghunathan2011}
\begin{barticle}[author]
\bauthor{\bsnm{He},~\bfnm{Yulei}\binits{Y.}},
  \bauthor{\bsnm{Yucel},~\bfnm{Recai}\binits{R.}} \AND
  \bauthor{\bsnm{Raghunathan},~\bfnm{Trivellore~E}\binits{T.~E.}}
(\byear{2011}).
\btitle{A functional multiple imputation approach to incomplete longitudinal
  data}.
\bjournal{Statistics in Medicine}
\bvolume{30}
\bpages{1137--1156}.
\bdoi{10.1002/sim.4201}
\end{barticle}
\endbibitem

\bibitem[\protect\citeauthoryear{Hirschman}{1994}]{Hirschman1994}
\begin{barticle}[author]
\bauthor{\bsnm{Hirschman},~\bfnm{Charles}\binits{C.}}
(\byear{1994}).
\btitle{Why Fertility Changes}.
\bjournal{Annual Review of Sociology}
\bvolume{20}
\bpages{203--233}.
\bdoi{10.1146/annurev.so.20.080194.001223}
\end{barticle}
\endbibitem

\bibitem[\protect\citeauthoryear{Honaker and King}{2010}]{HonakerKing2010}
\begin{barticle}[author]
\bauthor{\bsnm{Honaker},~\bfnm{James}\binits{J.}} \AND
  \bauthor{\bsnm{King},~\bfnm{Gary}\binits{G.}}
(\byear{2010}).
\btitle{What to Do about Missing Values in Time-Series Cross-Section Data}.
\bjournal{American Journal of Political Science}
\bvolume{54}
\bpages{561--581}.
\bdoi{10.1111/j.1540-5907.2010.00447.x}
\end{barticle}
\endbibitem

\bibitem[\protect\citeauthoryear{Ibrahim, Chen and Lipsitz}{2002}]{Ibrahim2002}
\begin{barticle}[author]
\bauthor{\bsnm{Ibrahim},~\bfnm{Joseph~G.}\binits{J.~G.}},
  \bauthor{\bsnm{Chen},~\bfnm{Ming-Hui}\binits{M.-H.}} \AND
  \bauthor{\bsnm{Lipsitz},~\bfnm{Stuart~R.}\binits{S.~R.}}
(\byear{2002}).
\btitle{Bayesian methods for generalized linear models with covariates missing
  at random}.
\bjournal{Canadian Journal of Statistics}
\bvolume{30}
\bpages{55--78}.
\bdoi{10.2307/3315865}
\end{barticle}
\endbibitem

\bibitem[\protect\citeauthoryear{Ibrahim, Lipsitz and Chen}{1999}]{Ibrahim1999}
\begin{barticle}[author]
\bauthor{\bsnm{Ibrahim},~\bfnm{Joseph~G}\binits{J.~G.}},
  \bauthor{\bsnm{Lipsitz},~\bfnm{Stuart~R}\binits{S.~R.}} \AND
  \bauthor{\bsnm{Chen},~\bfnm{M-H}\binits{M.-H.}}
(\byear{1999}).
\btitle{Missing covariates in generalized linear models when the missing data
  mechanism is non-ignorable}.
\bjournal{Journal of the Royal Statistical Society: Series B (Statistical
  Methodology)}
\bvolume{61}
\bpages{173--190}.
\bdoi{10.1111/1467-9868.00170}
\end{barticle}
\endbibitem

\bibitem[\protect\citeauthoryear{Kim et~al.}{2014}]{Kim2014}
\begin{barticle}[author]
\bauthor{\bsnm{Kim},~\bfnm{Hang~J.}\binits{H.~J.}},
  \bauthor{\bsnm{Reiter},~\bfnm{Jerome~P.}\binits{J.~P.}},
  \bauthor{\bsnm{Wang},~\bfnm{Quanli}\binits{Q.}},
  \bauthor{\bsnm{Cox},~\bfnm{Lawrence~H.}\binits{L.~H.}} \AND
  \bauthor{\bsnm{Karr},~\bfnm{Alan~F.}\binits{A.~F.}}
(\byear{2014}).
\btitle{Multiple Imputation of Missing or Faulty Values Under Linear
  Constraints}.
\bjournal{Journal of Business and Economic Statistics}
\bvolume{32}
\bpages{375--386}.
\bdoi{10.1080/07350015.2014.885435}
\end{barticle}
\endbibitem

\bibitem[\protect\citeauthoryear{King et~al.}{2001}]{King2001}
\begin{barticle}[author]
\bauthor{\bsnm{King},~\bfnm{Gary}\binits{G.}},
  \bauthor{\bsnm{Honaker},~\bfnm{James}\binits{J.}},
  \bauthor{\bsnm{Joseph},~\bfnm{Anne}\binits{A.}} \AND
  \bauthor{\bsnm{Scheve},~\bfnm{Kenneth}\binits{K.}}
(\byear{2001}).
\btitle{Analyzing Incomplete Political Science Data: An Alternative Algorithm
  for Multiple Imputation}.
\bjournal{American Political Science Review}
\bvolume{95}
\bpages{49--69}.
\bdoi{10.1017/S0003055401000235}
\end{barticle}
\endbibitem

\bibitem[\protect\citeauthoryear{Lee and Mitra}{2016}]{LeeMitra2016}
\begin{barticle}[author]
\bauthor{\bsnm{Lee},~\bfnm{Min~Cherng}\binits{M.~C.}} \AND
  \bauthor{\bsnm{Mitra},~\bfnm{Robin}\binits{R.}}
(\byear{2016}).
\btitle{Multiply imputing missing values in data sets with mixed measurement
  scales using a sequence of generalised linear models}.
\bjournal{Computational Statistics \& Data Analysis}
\bvolume{95}
\bpages{24--38}.
\bdoi{10.1016/j.csda.2015.08.004}
\end{barticle}
\endbibitem

\bibitem[\protect\citeauthoryear{Li, Yu and Rubin}{2012}]{LiYuRubin2012}
\begin{barticle}[author]
\bauthor{\bsnm{Li},~\bfnm{Fan}\binits{F.}},
  \bauthor{\bsnm{Yu},~\bfnm{Yaming}\binits{Y.}} \AND
  \bauthor{\bsnm{Rubin},~\bfnm{Donald~B.}\binits{D.~B.}}
(\byear{2012}).
\btitle{Imputing missing data by fully conditional models: Some cautionary
  examples and guidelines}.
\bjournal{Duke University Department of Statistical Science}.
\end{barticle}
\endbibitem

\bibitem[\protect\citeauthoryear{Lipsitz and
  Ibrahim}{1996}]{LipsitzIbrahim1996}
\begin{barticle}[author]
\bauthor{\bsnm{Lipsitz},~\bfnm{Stuart~R.}\binits{S.~R.}} \AND
  \bauthor{\bsnm{Ibrahim},~\bfnm{Joseph~G.}\binits{J.~G.}}
(\byear{1996}).
\btitle{A conditional model for incomplete covariates in parametric regression
  models}.
\bjournal{Biometrika}
\bvolume{83}
\bpages{916--922}.
\bdoi{10.1093/biomet/83.4.916}
\end{barticle}
\endbibitem

\bibitem[\protect\citeauthoryear{Little and Rubin}{2002}]{LittleRubin2002}
\begin{bbook}[author]
\bauthor{\bsnm{Little},~\bfnm{Roderick J.~A.}\binits{R.~J.~A.}} \AND
  \bauthor{\bsnm{Rubin},~\bfnm{Donald~B.}\binits{D.~B.}}
(\byear{2002}).
\btitle{Statistical Analysis with Missing Data}.
\bpublisher{John Wiley \& Sons Inc.}
\end{bbook}
\endbibitem

\bibitem[\protect\citeauthoryear{Liu and Raftery}{2025}]{LiuRaftery2025}
\begin{barticle}[author]
\bauthor{\bsnm{Liu},~\bfnm{Daphne~H.}\binits{D.~H.}} \AND
  \bauthor{\bsnm{Raftery},~\bfnm{Adrian~E.}\binits{A.~E.}}
(\byear{2025}).
\btitle{Supplement to ``Multiple Imputation of Hierarchical
  Nonlinear Time Series Data with an Application to School Enrollment Data''}.
\end{barticle}
\endbibitem

\bibitem[\protect\citeauthoryear{Liu, Taylor and
  Belin}{2000}]{LiuTaylorBelin2000}
\begin{barticle}[author]
\bauthor{\bsnm{Liu},~\bfnm{Minzhi}\binits{M.}},
  \bauthor{\bsnm{Taylor},~\bfnm{Jeremy M.~G.}\binits{J.~M.~G.}} \AND
  \bauthor{\bsnm{Belin},~\bfnm{Thomas~R.}\binits{T.~R.}}
(\byear{2000}).
\btitle{Multiple Imputation and Posterior Simulation for Multivariate Missing
  Data in Longitudinal Studies}.
\bjournal{Biometrics}
\bvolume{56}
\bpages{1157--1163}.
\bdoi{10.1111/j.0006-341X.2000.01157.x}
\end{barticle}
\endbibitem

\bibitem[\protect\citeauthoryear{L{\"u}dtke, Robitzsch and
  Grund}{2017}]{LudtkeRobitzschGrund2017}
\begin{barticle}[author]
\bauthor{\bsnm{L{\"u}dtke},~\bfnm{Oliver}\binits{O.}},
  \bauthor{\bsnm{Robitzsch},~\bfnm{Alexander}\binits{A.}} \AND
  \bauthor{\bsnm{Grund},~\bfnm{Simon}\binits{S.}}
(\byear{2017}).
\btitle{Multiple Imputation of Missing Data in Multilevel Designs: A Comparison
  of Different Strategies}.
\bjournal{Psychological Methods}
\bvolume{22}
\bpages{141--165}.
\bdoi{10.1037/met0000096}
\end{barticle}
\endbibitem

\bibitem[\protect\citeauthoryear{L{\"u}dtke, Robitzsch and
  West}{2020}]{Ludtke2020}
\begin{barticle}[author]
\bauthor{\bsnm{L{\"u}dtke},~\bfnm{Oliver}\binits{O.}},
  \bauthor{\bsnm{Robitzsch},~\bfnm{Alexander}\binits{A.}} \AND
  \bauthor{\bsnm{West},~\bfnm{Stephen~G.}\binits{S.~G.}}
(\byear{2020}).
\btitle{Regression Models Involving Nonlinear Effects With Missing Data: A
  Sequential Modeling Approach Using Bayesian Estimation}.
\bjournal{Psychological Methods}
\bvolume{25}
\bpages{157--181}.
\bdoi{10.1037/met0000233}
\end{barticle}
\endbibitem

\bibitem[\protect\citeauthoryear{Mbougua et~al.}{2013}]{Mbougua2013}
\begin{barticle}[author]
\bauthor{\bsnm{Mbougua},~\bfnm{Jules Brice~Tchatchueng}\binits{J.~B.~T.}},
  \bauthor{\bsnm{Laurent},~\bfnm{Christian}\binits{C.}},
  \bauthor{\bsnm{Ndoye},~\bfnm{ibra}\binits{i.}},
  \bauthor{\bsnm{Delaporte},~\bfnm{Eric}\binits{E.}},
  \bauthor{\bsnm{Gwetb},~\bfnm{Henri}\binits{H.}} \AND
  \bauthor{\bsnm{Molinarie},~\bfnm{Nicolas}\binits{N.}}
(\byear{2013}).
\btitle{Nonlinear multiple imputation for continuous covariate within
  semiparametric Cox model: Application to HIV data in Senegal}.
\bjournal{Statistics in Medicine}
\bvolume{32}
\bpages{4651--4665}.
\bdoi{10.1002/sim.5854}
\end{barticle}
\endbibitem

\bibitem[\protect\citeauthoryear{Meng}{1994}]{Meng1994}
\begin{barticle}[author]
\bauthor{\bsnm{Meng},~\bfnm{Xiao-Li}\binits{X.-L.}}
(\byear{1994}).
\btitle{Multiple-Imputation Inferences with Uncongenial Sources of Input}.
\bjournal{Statistical Science}
\bvolume{9}
\bpages{538--558}.
\bdoi{10.1214/ss/1177010269}
\end{barticle}
\endbibitem

\bibitem[\protect\citeauthoryear{Nguyen, Carlin and Lee}{2017}]{Nguyen2017}
\begin{barticle}[author]
\bauthor{\bsnm{Nguyen},~\bfnm{Cattram~D.}\binits{C.~D.}},
  \bauthor{\bsnm{Carlin},~\bfnm{John~B.}\binits{J.~B.}} \AND
  \bauthor{\bsnm{Lee},~\bfnm{Katherine~J.}\binits{K.~J.}}
(\byear{2017}).
\btitle{Model checking in multiple imputation: an overview and case study}.
\bjournal{Emerging Themes in Epidemiology}
\bvolume{14}.
\bdoi{10.1186/s12982-017-0062-6}
\end{barticle}
\endbibitem

\bibitem[\protect\citeauthoryear{Raftery and Lewis}{1996}]{RafteryLewis1996}
\begin{bincollection}[author]
\bauthor{\bsnm{Raftery},~\bfnm{Adrian~E.}\binits{A.~E.}} \AND
  \bauthor{\bsnm{Lewis},~\bfnm{Steven~M.}\binits{S.~M.}}
(\byear{1996}).
\btitle{Implementing {MCMC}}.
In \bbooktitle{Markov Chain Monte Carlo in Practice}
(\beditor{\bfnm{W.~R.}\binits{W.~R.}~\bsnm{Gilks}},
  \beditor{\bfnm{S.}\binits{S.}~\bsnm{Richardson}} \AND
  \beditor{\bfnm{David}\binits{D.}~\bsnm{Spiegelhalter}}, eds.)
\bpages{115--130}.
\bpublisher{Chapman and Hall}, \baddress{London}.
\end{bincollection}
\endbibitem

\bibitem[\protect\citeauthoryear{Raghunathan et~al.}{2001}]{Raghunathan2001}
\begin{barticle}[author]
\bauthor{\bsnm{Raghunathan},~\bfnm{Trivellore~E.}\binits{T.~E.}},
  \bauthor{\bsnm{Lepkowski},~\bfnm{James~M.}\binits{J.~M.}},
  \bauthor{\bsnm{Van~Hoewyk},~\bfnm{John}\binits{J.}} \AND
  \bauthor{\bsnm{Solenberger},~\bfnm{Peter}\binits{P.}}
(\byear{2001}).
\btitle{A Multivariate Technique for Multiply Imputing Missing Values Using a
  Sequence of Regression Models}.
\bjournal{Survey Methodology}
\bvolume{27}
\bpages{85--95}.
\end{barticle}
\endbibitem

\bibitem[\protect\citeauthoryear{Rubin}{1976}]{Rubin1976}
\begin{barticle}[author]
\bauthor{\bsnm{Rubin},~\bfnm{Donald~B.}\binits{D.~B.}}
(\byear{1976}).
\btitle{Inference and Missing Data}.
\bjournal{Biometrika}
\bvolume{63}
\bpages{581--592}.
\bdoi{10.1093/biomet/63.3.581}
\end{barticle}
\endbibitem

\bibitem[\protect\citeauthoryear{Rubin}{1977}]{Rubin1977}
\begin{barticle}[author]
\bauthor{\bsnm{Rubin},~\bfnm{Donald~B.}\binits{D.~B.}}
(\byear{1977}).
\btitle{Formalizing Subjective Notions About the Effect of Nonrespondents in
  Sample Surveys}.
\bjournal{Journal of the American Statistical Association}
\bvolume{72}
\bpages{538--543}.
\bdoi{10.1080/01621459.1977.10480610}
\end{barticle}
\endbibitem

\bibitem[\protect\citeauthoryear{Rubin}{1978}]{Rubin1978}
\begin{binproceedings}[author]
\bauthor{\bsnm{Rubin},~\bfnm{Donald~B.}\binits{D.~B.}}
(\byear{1978}).
\btitle{Multiple Imputations in Sample Surveys-A Phenomenological Bayesian
  Approach to Nonresponse}.
In \bbooktitle{Proceedings of the Survey Research Methods Section of the
  American Statistical Association}
\bvolume{1}
\bpages{20--34}.
\bpublisher{American Statistical Association}.
\end{binproceedings}
\endbibitem

\bibitem[\protect\citeauthoryear{Rubin}{1987}]{Rubin1987}
\begin{bbook}[author]
\bauthor{\bsnm{Rubin},~\bfnm{Donald~B.}\binits{D.~B.}}
(\byear{1987}).
\btitle{Multiple Imputation for Nonresponse in Surveys}.
\bpublisher{John Wiley \& Sons Inc.}
\end{bbook}
\endbibitem

\bibitem[\protect\citeauthoryear{Rubin}{1996}]{Rubin1996}
\begin{barticle}[author]
\bauthor{\bsnm{Rubin},~\bfnm{Donald~B.}\binits{D.~B.}}
(\byear{1996}).
\btitle{Multiple Imputation After 18+ Years}.
\bjournal{Journal of the American Statistical Association}
\bvolume{91}
\bpages{473--489}.
\end{barticle}
\endbibitem

\bibitem[\protect\citeauthoryear{Rubin and Schafer}{1990}]{RubinSchafer1990}
\begin{binproceedings}[author]
\bauthor{\bsnm{Rubin},~\bfnm{Donald~B.}\binits{D.~B.}} \AND
  \bauthor{\bsnm{Schafer},~\bfnm{Joseph~L.}\binits{J.~L.}}
(\byear{1990}).
\btitle{Efficiently creating multiple imputations for incomplete multivariate
  normal data}.
In \bbooktitle{Proceedings of the Statistical Computing Section of the American
  Statistical Association}
\bpages{83-88}.
\bpublisher{American Statistical Association}.
\end{binproceedings}
\endbibitem

\bibitem[\protect\citeauthoryear{Savalei and
  Rhemtulla}{2012}]{SavaleiRhemtulla2012}
\begin{barticle}[author]
\bauthor{\bsnm{Savalei},~\bfnm{Victoria}\binits{V.}} \AND
  \bauthor{\bsnm{Rhemtulla},~\bfnm{Mijke}\binits{M.}}
(\byear{2012}).
\btitle{On Obtaining Estimates of the Fraction of Missing Information From Full
  Information Maximum Likelihood}.
\bjournal{Structural Equation Modeling: A Multidisciplinary Journal}
\bvolume{19}
\bpages{477--494}.
\bdoi{10.1080/10705511.2012.687669}
\end{barticle}
\endbibitem

\bibitem[\protect\citeauthoryear{Schafer}{1997a}]{Schafer1997}
\begin{bbook}[author]
\bauthor{\bsnm{Schafer},~\bfnm{Joseph~L.}\binits{J.~L.}}
(\byear{1997}a).
\btitle{Analysis of Incomplete Multivariate Data}.
\bpublisher{CRC Press}.
\end{bbook}
\endbibitem

\bibitem[\protect\citeauthoryear{Schafer}{1997b}]{Schafer1997TechnicalReport}
\begin{btechreport}[author]
\bauthor{\bsnm{Schafer},~\bfnm{Joseph~L.}\binits{J.~L.}}
(\byear{1997}b).
\btitle{Imputation of missing covariates under a multivariate linear mixed
  model.}
\btype{Technical Report},
\bpublisher{Dept. of Statistics, The Pennsylvania State University}.
\end{btechreport}
\endbibitem

\bibitem[\protect\citeauthoryear{Schafer and Olsen}{1998}]{SchaferOlsen1998}
\begin{barticle}[author]
\bauthor{\bsnm{Schafer},~\bfnm{Joseph~L.}\binits{J.~L.}} \AND
  \bauthor{\bsnm{Olsen},~\bfnm{Maren~K.}\binits{M.~K.}}
(\byear{1998}).
\btitle{Multiple Imputation for Multivariate Missing-Data Problems: A Data
  Analyst's Perspective}.
\bjournal{Multivariate Behavioral Research}
\bvolume{33}
\bpages{545--571}.
\bdoi{10.1207/s15327906mbr3304_5}
\end{barticle}
\endbibitem

\bibitem[\protect\citeauthoryear{Schafer and Yucel}{2002}]{SchaferYucel2002}
\begin{barticle}[author]
\bauthor{\bsnm{Schafer},~\bfnm{Joseph~L.}\binits{J.~L.}} \AND
  \bauthor{\bsnm{Yucel},~\bfnm{Recai~M.}\binits{R.~M.}}
(\byear{2002}).
\btitle{Computational Strategies for Multivariate Linear Mixed-Effects Models
  With Missing Values}.
\bjournal{Journal of Computational and Graphical Statistics}
\bvolume{11}
\bpages{437--457}.
\bdoi{10.1198/106186002760180608}
\end{barticle}
\endbibitem

\bibitem[\protect\citeauthoryear{Speidel, Drechsler and
  Jolani}{2018}]{Speidel2018}
\begin{btechreport}[author]
\bauthor{\bsnm{Speidel},~\bfnm{Matthias}\binits{M.}},
  \bauthor{\bsnm{Drechsler},~\bfnm{J{\"o}rg}\binits{J.}} \AND
  \bauthor{\bsnm{Jolani},~\bfnm{Shahab}\binits{S.}}
(\byear{2018}).
\btitle{R package hmi: A convenient tool for hierarchical multiple imputation
  and beyond}
\btype{Technical Report} No. \bnumber{16},
\bpublisher{IAB-Discussion Paper}.
\end{btechreport}
\endbibitem

\bibitem[\protect\citeauthoryear{Taljaard, Donner and
  Klar}{2008}]{TaljaardDonnerKlar2008}
\begin{barticle}[author]
\bauthor{\bsnm{Taljaard},~\bfnm{Monica}\binits{M.}},
  \bauthor{\bsnm{Donner},~\bfnm{Allan}\binits{A.}} \AND
  \bauthor{\bsnm{Klar},~\bfnm{Neil}\binits{N.}}
(\byear{2008}).
\btitle{Imputation Strategies for Missing Continuous Outcomes in Cluster
  Randomized Trials}.
\bjournal{Biometrical Journal}
\bvolume{50}
\bpages{329--345}.
\bdoi{10.1002/bimj.200710423}
\end{barticle}
\endbibitem

\bibitem[\protect\citeauthoryear{Tanner and Wong}{1987}]{TannerWong1987}
\begin{barticle}[author]
\bauthor{\bsnm{Tanner},~\bfnm{Martin~A.}\binits{M.~A.}} \AND
  \bauthor{\bsnm{Wong},~\bfnm{Wing~Hung}\binits{W.~H.}}
(\byear{1987}).
\btitle{The Calculation of Posterior Distributions by Data Augmentation}.
\bjournal{Journal of the American Statistical Association}
\bvolume{82}
\bpages{528--540}.
\bdoi{10.1080/01621459.1987.10478458}
\end{barticle}
\endbibitem

\bibitem[\protect\citeauthoryear{{UNESCO Institute for
  Statistics}}{2023}]{UNESCO}
\begin{bmisc}[author]
\bauthor{\bsnm{{UNESCO Institute for Statistics}}}
(\byear{2023}).
\btitle{Background Information on Education Statistics in the {UIS} Database}.
\end{bmisc}
\endbibitem


\bibitem[\protect\citeauthoryear{{United Nations, Department of Economic and
  Social Affairs, Population Division}}{2022}]{WPP2022}
\begin{bmisc}[author]
\bauthor{\bsnm{{United Nations, Department of Economic and Social Affairs,
  Population Division}}}
(\byear{2022}).
\btitle{World Population Prospects 2022}.
\bhowpublished{Online Edition}.
\end{bmisc}
\endbibitem

\bibitem[\protect\citeauthoryear{van Buuren and
  Groothuis-Oudshoorn}{2011}]{vanBuurenOudshoorn2011}
\begin{barticle}[author]
\bauthor{\bparticle{van} \bsnm{Buuren},~\bfnm{Stef}\binits{S.}} \AND
  \bauthor{\bsnm{Groothuis-Oudshoorn},~\bfnm{Karin}\binits{K.}}
(\byear{2011}).
\btitle{{mice}: Multivariate Imputation by Chained Equations in R}.
\bjournal{Journal of Statistical Software}
\bvolume{45}
\bpages{1--67}.
\bdoi{10.18637/jss.v045.i03}
\end{barticle}
\endbibitem

\bibitem[\protect\citeauthoryear{van Buuren et~al.}{2006}]{vanBuuren2006}
\begin{barticle}[author]
\bauthor{\bparticle{van} \bsnm{Buuren},~\bfnm{Stef}\binits{S.}},
  \bauthor{\bsnm{Brand},~\bfnm{Jaap~PL}\binits{J.~P.}},
  \bauthor{\bsnm{Groothuis-Oudshoorn},~\bfnm{Catharina~GM}\binits{C.~G.}} \AND
  \bauthor{\bsnm{Rubin},~\bfnm{Donald~B}\binits{D.~B.}}
(\byear{2006}).
\btitle{Fully conditional specification in multivariate imputation}.
\bjournal{Journal of Statistical Computation and Simulation}
\bvolume{76}
\bpages{1049--1064}.
\bdoi{10.1080/10629360600810434}
\end{barticle}
\endbibitem

\bibitem[\protect\citeauthoryear{{World Bank}}{2021}]{WorldBank}
\begin{bmisc}[author]
\bauthor{\bsnm{{World Bank}}}
(\byear{2021}).
\btitle{World Bank Open Data: School enrollment, secondary (\% gross and \%
  net)}.
\bhowpublished{Available at: \url{https://data.worldbank.org/}}.
\end{bmisc}
\endbibitem

\bibitem[\protect\citeauthoryear{Xie and Meng}{2017}]{XieMeng2017}
\begin{barticle}[author]
\bauthor{\bsnm{Xie},~\bfnm{Xianchao}\binits{X.}} \AND
  \bauthor{\bsnm{Meng},~\bfnm{Xiao-Li}\binits{X.-L.}}
(\byear{2017}).
\btitle{Dissecting Multiple Imputation From a Multi-Phase Inference
  Perspective: What Happens When God's, Imputer's and Analyst's Models are
  Uncongenial?}
\bjournal{Statistica Sinica}
\bvolume{27}
\bpages{1485--1545}.
\bdoi{10.5705/ss.2014.067}
\end{barticle}
\endbibitem

\bibitem[\protect\citeauthoryear{Xu, Daniels and Winterstein}{2016}]{Xu2016}
\begin{barticle}[author]
\bauthor{\bsnm{Xu},~\bfnm{Dandan}\binits{D.}},
  \bauthor{\bsnm{Daniels},~\bfnm{Michael~J.}\binits{M.~J.}} \AND
  \bauthor{\bsnm{Winterstein},~\bfnm{Almut~G.}\binits{A.~G.}}
(\byear{2016}).
\btitle{Sequential BART for imputation of missing covariates}.
\bjournal{Biostatistics}
\bvolume{17}
\bpages{589--602}.
\bdoi{10.1093/biostatistics/kxw009}
\end{barticle}
\endbibitem

\bibitem[\protect\citeauthoryear{Zhao and Schafer}{2023}]{pan}
\begin{bmisc}[author]
\bauthor{\bsnm{Zhao},~\bfnm{Jing~Hua}\binits{J.~H.}} \AND
  \bauthor{\bsnm{Schafer},~\bfnm{Joseph~L.}\binits{J.~L.}}
(\byear{2023}).
\btitle{pan: Multiple imputation for multivariate panel or clustered data}.
\bnote{R package version 1.8}.
\end{bmisc}
\endbibitem

\end{thebibliography}


\end{document}